\documentclass[aps,superscriptaddress,twocolumn,dvips]{revtex4}
\usepackage{feynmp}
\usepackage{amsmath}
\usepackage{amssymb}
\usepackage{epsfig}
\usepackage{graphicx}
\usepackage{subfigure}
\usepackage{hyperref}
\usepackage{bbm}
\usepackage{color}
\usepackage{longtable}
\usepackage{booktabs}
\usepackage{multirow}

\newcommand{\Rmnum}[1]{\expandafter\@slowromancap\romannumeral #1@}
\allowdisplaybreaks[3]
\begin{document}

\title{Quantum Criticality of Excitonic Insulating Transition in Nodal Line
Semimetal ZrSiS}

\author{Jing-Rong Wang}
\affiliation{Anhui Province Key Laboratory of Condensed Matter
Physics at Extreme Conditions, High Magnetic Field Laboratory of the
Chinese Academy of Sciences, Hefei 230031, China}
\author{Guo-Zhu Liu}
\altaffiliation{Corresponding author: gzliu@ustc.edu.cn}
\affiliation{Department of Modern Physics, University of Science and
Technology of China, Hefei 230026, China}
\author{Xiangang Wan}
\affiliation{National Laboratory of Solid State Microstructures,
College of Physics, Nanjing University, Nanjing 210093, China}
\affiliation{Collaborative Innovation Center of Advanced
Microstructures, Nanjing University, Nanjing 210093, China}
\author{Changjin Zhang}
\altaffiliation{Corresponding author: zhangcj@hmfl.ac.cn}
\affiliation{Anhui Province Key Laboratory of Condensed Matter
Physics at Extreme Conditions, High Magnetic Field Laboratory of the
Chinese Academy of Sciences, Hefei 230031, China}
\affiliation{Collaborative Innovation Center of Advanced
Microstructures, Nanjing University, Nanjing 210093, China}
\affiliation{Institutes of Physical Science and Information
Technology, Anhui University, Hefei 230601, China}

\begin{abstract}
Pezzini \emph{et al.} reported an unconventional mass enhancement in
topological nodal line semimetal ZrSiS (Nat. Phys. {\bf 14}, 178
(2018), whose origin remains puzzling. In this material, strong
short-range interactions might induce excitonic particle-hole pairs.
Here we study the renormalization of fermion velocities and find
that the mass enhancement in ZrSiS can be well understood if we
suppose that ZrSiS is close to the quantum critical point between
semimetal and excitonic insulator. Near this quantum critical point,
the fermion velocities are considerably reduced by excitonic quantum
fluctuation, leading to fermion mass enhancement. The quasiparticle
residue is suppressed as the energy decreases but is finite at zero
energy. This indicates that ZrSiS is a strongly correlated Fermi
liquid, and explains why the mass enhancement is weaker than
non-Fermi liquids. Our results suggest that ZrSiS is a rare example
of 3D topological semimetal exhibiting unusual quantum criticality.
\end{abstract}

\maketitle


Quantum phase transition has emerged as one of the most important
subjects in modern condensed matter physics. For a second-order
transition, thermodynamic phases with distinct symmetries are
separated by a quantum critical point (QCP). In the vicinity of the
QCP, the low-energy quasiparticle properties are governed by the
quantum fluctuation of order parameter such that some interesting
unconventional phenomena could arise, e.g., a significant
enhancement or divergence of quasiparticle mass \cite{Rosch07}. In
real systems, quantum criticality exhibiting greatly enhanced mass
has been observed in a variety of materials, including cuprates
YBa$_2$Cu$_3$O$_{6+\delta}$ \cite{Ramshaw15} and Eu(Nd)-LSCO
\cite{Taillefer19}, heavy fermion compounds YbRh$_2$Si$_2$,
CeCoIn$_5$, and CeRhIn$_5$ \cite{Gegenwart02, Tokiwa13, Aynajian12},
iron pnictide BaFe(As$_x$P$_{1-x}$)$_2$ \cite{Walmsley13}, and
spinel magnet ZnCr$_2$Se$_4$ \cite{GuCC18}.

In the past decade, much efforts have been put into the topological
semimetal (SM) materials \cite{Wan11, XuG11, LiuZK14, Yan17,
LiCequn18, Armitage18, Wan19, FangC19, Vergniory19}. In these SMs,
most behaviors are manifested by their single-particle characters.
However, under certain conditions, inter-particle interactions can
markedly renormalize the energy dispersion of free fermions, and
even cause quantum phase transitions \cite{Gonzalez99, Kotov12,
Hofmann14, Elias11, Goawami11, Hosur12, Moon13, Isobe16, Hirata16,
LiuD16, Hirata17, Roy17}. For SMs tuned close to band-touching
point, there is perfect particle-hole symmetry and the semimetallic
ground state may become instable due to particle-hole pairing
\cite{Keldysh}, leading to an excitonic insulating transition. This
would be a nice platform to study quantum critical phenomena in SMs,
such as unusual mass enhancement. Nevertheless, excitonic
instability requires a sufficiently strong long-range Coulomb
interaction (LRCI) or short-range four-fermion interaction
\cite{Drut09, WeiH12, WangLiuZhang17A, Rudenko18, Roy18}, which is
not easy to realize in actual systems. Although there are many
proposals for excitonic transition in various SMs \cite{Drut09,
WeiH12, WangLiuZhang17A, Rudenko18, Roy18, Herbut14}, so far
experimental evidences of excitonic QCP and the associated quantum
critical phenomena are still limited.

Recently, Pezzini \emph{et al.} \cite{Pezzini18} have measured the
quantum oscillations of ZrSiS, a topological nodal line SM (NLSM),
and revealed an unconventional enhancement of quasiparticle mass.
Similar to what happens in cuprate YBa$_2$Cu$_3$O$_{6+\delta}$ and
other materials, the observed mass enhancement is conjectured
\cite{Pezzini18} to be induced by electronic correlations. The
physical mechanism underlying this observation is still
undetermined. Numerical calculations based on a simplified
two-dimensional model \cite{Rudenko18} suggested that ZrSiS may
become an excitonic insulator at low temperatures due to strong
short-range interaction, which opens a finite pseudogap in the
electronic spectrum. However, the opening of a finite excitonic gap
is in contradiction to angle-resolved photoemission spectroscopy
(ARPES) \cite{Schoop16, Neupane16, FuBB19} and transport
\cite{Ali16, Singha17} experiments on ZrSiS. In order to resolve
this puzzle, one should search for a physical mechanism that
enhances the quasiparticle mass and meanwhile is not at odds with
the existing ARPES and transport experiments. It is also interesting
to understand why the mass enhancement of ZrSiS is not as
significant as the previously mentioned unconventional
superconductors \cite{Ramshaw15, Taillefer19, Gegenwart02, Tokiwa13,
Aynajian12, Walmsley13}.

\begin{figure}[htbp]
\center
\includegraphics[width=2.66in]{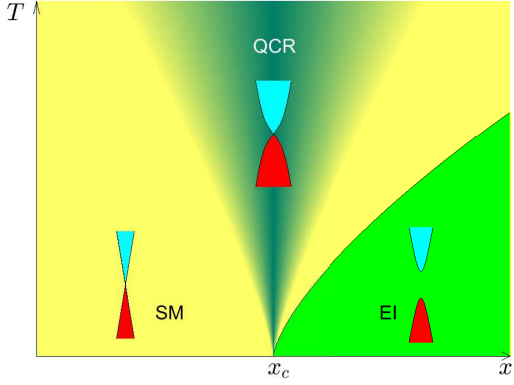}
\caption{Schematic global phase diagram of NLSM. Here, SM stands for
semimetal, EI for excitonic insulator, and QCR for quantum critical
region. SM-EI transition occurs by tuning a non-thermal parameter
$x$, and $x_c$ is the QCP. In QCR, the excitonic quantum fluctuation
leads to considerable fermion mass enhancement.
\label{Fig:PhaseDiagram}}
\end{figure}

In this Letter, we analyze the dynamics of nodal line fermions near
an excitonic instability of NLSM and demonstrate that the excitonic
quantum fluctuation can considerably increase the fermion mass
without gapping the system. This provides a promising explanation of
the unconventional mass enhancement of ZrSiS \cite{Pezzini18}. To
illustrate our proposal, we plot a schematic phase diagram in
Fig.~\ref{Fig:PhaseDiagram}. Tuning the parameter $x$ drives a
quantum phase transition between NLSM and excitonic insulator (EI),
with $x_c$ being the QCP. The system is gapless in SM phase ($x <
x_c$), and acquires a gap in EI phase due to excitonic pairing ($x >
x_c$). For ZrSiS, $x$ can be identified as the strength parameter of
short-range interaction. First of all, we infer that ZrSiS cannot be
deep in the EI phase, since experiments \cite{Schoop16, Neupane16,
FuBB19, Ali16, Singha17} did not observe a finite gap. Deep in the
SM phase, there are no excitonic pairs, and nodal line fermions are
subjected to LRCI and weak short-range interaction. RG analysis show
that, LRCI tends to reduce the mass, whereas the weak short-range
interaction does not renormalize the mass. Thus, ZrSiS should not be
deep in the SM phase. Our proposal is that, the short-range
interaction in ZrSiS is not strong enough to turn it into the EI
phase, but suffices to drive this material to fall into the
intermediate quantum critical region (QCR). In this QCR, the
excitonic order parameter has a vanishing mean value and the nodal
line fermions remain gapless, in accordance with experiments
\cite{Schoop16, Neupane16, FuBB19}. However, different from SM
phase, the quantum fluctuations of excitonic pairs are important in
QCR and can substantially renormalize the quasiparticle mass.

To study what could happen if ZrSiS is in the QCR, we carefully
treat the interaction between nodal line fermions and excitonic
fluctuation by using the renormalization group (RG) approach, and
find that the fermion velocities are suppressed, which enhances the
effective fermion mass. The quasiparticle residue $Z_f$ decreases
rapidly as the energy is lowered, but flows to a small finite value
in the zero-energy limit. Thus, ZrSiS is a strongly interacting
Fermi liquid near the SM-EI QCP. This explains why the effective
mass in ZrSiS is considerably enhanced but does not take a large
value.

\emph{Model.} The action of the free nodal line fermions is
\begin{eqnarray}
S_{\psi} = \int\frac{d\omega}{2\pi} \frac{d^{3}
\mathbf{k}}{(2\pi)^{3}} \psi_{a}^{\dag}(\omega,\mathbf{k})
\left(-i\omega+\mathcal{H}_{0}(\mathbf{k}) \right)
\psi_{a}(\omega,\mathbf{k}),
\end{eqnarray}
where Hamiltonian $\mathcal{H}_{0}(\mathbf{k}) =
\frac{\left(k_{\bot}^{2} - k_{F}^{2}\right)}{2m}\sigma_{1} +
v_{z}k_{z}\sigma_{2}$ \cite{Huh16, Roy17}. Here, $k_{\bot}^{2} =
k_{x}^2 + k_{y}^2$. Near the nodal lines, one can approximate
$\mathcal{H}_{0}$ by $\mathcal{H}_{0}\approx
v_{F}k_{r}\sigma_{1}+v_{z}k_{z}\sigma_{2}$, where $k_{r} =
k_{\bot}-k_{F}$, $v_{F}=k_{F}/m$ is the fermion velocity within the
$x$-$y$ plane and $v_{z}$ is fermion velocity along $z$-axis. The
interaction-induced renormalization of fermion mass can be easily
obtained from the renormalization of velocities $v_F$ and $v_z$.
$\psi_{a}$ is a two component spinor field, and $\sigma_{1,2,3}$ are
Pauli matrices. The index $a=1,2,...,N$ where $N$ is the fermion
flavor. In NLSM, the density of states (DOS) behaves as
$\rho(\omega)\propto\omega$, which vanishes at the Fermi level ,
i.e. $\rho(0) = 0$. Notice that, in the simplified two-dimensional
model studied in Ref.~\cite{Rudenko18}, $\rho(0)$ takes a large
finite value. This is because the dispersion of nodal line fermions
along $z$-axis is completely neglected in that model.

$\mathcal{H}_{0}$ respects the chiral symmetry:
$\left\{\mathcal{H}_{0},\sigma_{3}\right\} = 0$. If the fermion
acquires a mass term $\mathcal{H}_{\Delta} = \Delta\sigma_{3}$ due
to excitonic pairing, a finite gap is opened, breaking the chiral
symmetry, and the system becomes an EI (amounting to a charge
density wave).

The action of excitonic quantum fluctuation, represented by scalar
field $\phi$, takes the form
\begin{eqnarray}
S_{\phi} = \int\frac{d\Omega}{2\pi}\frac{d^{3}
\mathbf{q}}{(2\pi)^{3}}\phi(\Omega,\mathbf{q}) \left(\Omega^{2} +
E_{b}^{2} + r\right)\phi(\Omega,\mathbf{k}),
\end{eqnarray}
where $E_{b}^{2} = v_{b\bot}^{2}q_{\bot}^{2}+v_{bz}^{2} q_{z}^{2}$.
Here, $v_{b\bot}$ and $v_{bz}$ are boson velocities within $x$-$y$
plane and along $z$-axis. Boson mass $r \propto (x-x_c)$ measures
the distance of the system to SM-EI QCP, and $r=0$ at QCP. The
Yukawa coupling between fermion and boson is described by
\begin{eqnarray}
S_{\psi\phi} = g\int d\tau d^3\mathbf{x}
\psi_{a}^{\dag}(\tau,\mathbf{x})\sigma_{3}
\psi_{a}(\tau,\mathbf{x})\phi(\tau,\mathbf{x}).
\end{eqnarray}
There is also a $\phi^{4}$ interaction: $
S_{\phi^{4}}=\frac{u}{24}\int d\tau d^3\mathbf{x}
\phi^4(\tau,\mathbf{x})$. Here, $g$ and $u$ are coupling
coefficients.

\emph{Renormalization group results.} The model parameters appearing
in the total action $S_{\psi} + S_{\phi} + S_{\psi\phi} +
S_{\phi^4}$ are renormalized by interactions. Their RG equations,
derived in the Supplementary Materials, are given by
\begin{eqnarray}
\frac{dv_{F}}{d\ell}&=&\left(C_{1}-C_{0}\right)v_{F},
\\
\frac{dv_{z}}{d\ell}&=&\left(C_{2}-C_{0}\right)v_{z},
\\
\frac{dv_{b\bot}}{d\ell}&=& \frac{1}{2} \left(C_{\bot} - C_{\phi}
\right) v_{b\bot},
\\
\frac{dv_{bz}}{d\ell}&=&\frac{1}{2}\left(C_{z}-C_{\phi}\right)
v_{bz},
\\
\frac{dZ_{f}}{d\ell}&=&-C_{0}Z_{f},
\\
\frac{d\alpha_{g}}{d\ell} &=& -\left(-C_{0} + 3C_{1} + 2C_{3} +
C_{\phi}\right)\alpha_{g}, \\
\frac{d\beta_{g}}{d\ell} &=& \left(1-C_{\phi}-2C_{1} -
2C_{3}\right)\beta_{g},
\\
\frac{du}{d\ell} &=& -\left(C_{\bot}+\frac{C_{z}}{2} +
\frac{C_{\phi}}{2}\right)u+\frac{3u^{2}}{16}+\frac{12\alpha_{g}
\beta_{g}}{\pi \delta_{1}\delta_{2}^{2}\delta_{3}},
\end{eqnarray}
where $\ell$ is a running parameter. $\alpha_{g}=g^{2}/v_{F}^{3}$ is
a dimensionless coupling parameter for the Yukawa coupling, and
$\beta_{g} = \frac{Ng^{2}k_{F}}{8\pi v_{F}^{2}\Lambda}$. We
defined three parameters: $\delta_{1} = v_{z}/v_{F}$, $\delta_{2} =
v_{b\bot}/v_{F}$, and $\delta_{3} = v_{bz}/v_{F}$. The expressions
for parameters $C_{i}\equiv C_{i}(\alpha_{g},\delta_{1},
\delta_{2},\delta_{3})$, where $i = 0,1,2,3$, are given in
Supplementary Materials. We also define
$C_{\phi}=\beta_{g}/\delta_{1}$,
$C_{\bot}=\frac{\beta_{g}}{4\delta_{1}\delta_{2}^{2}}$, and
$C_{z}=\frac{\beta_{g}\delta_{1}}{2\delta_{3}^{2}}$. $Z_f$ is the
quasiparticle residue: $Z_f \neq 0$ for Fermi liquids and $Z_f = 0$
for non-Fermi liquids.

\begin{figure}[htbp]
\center
\includegraphics[width=3.3in]{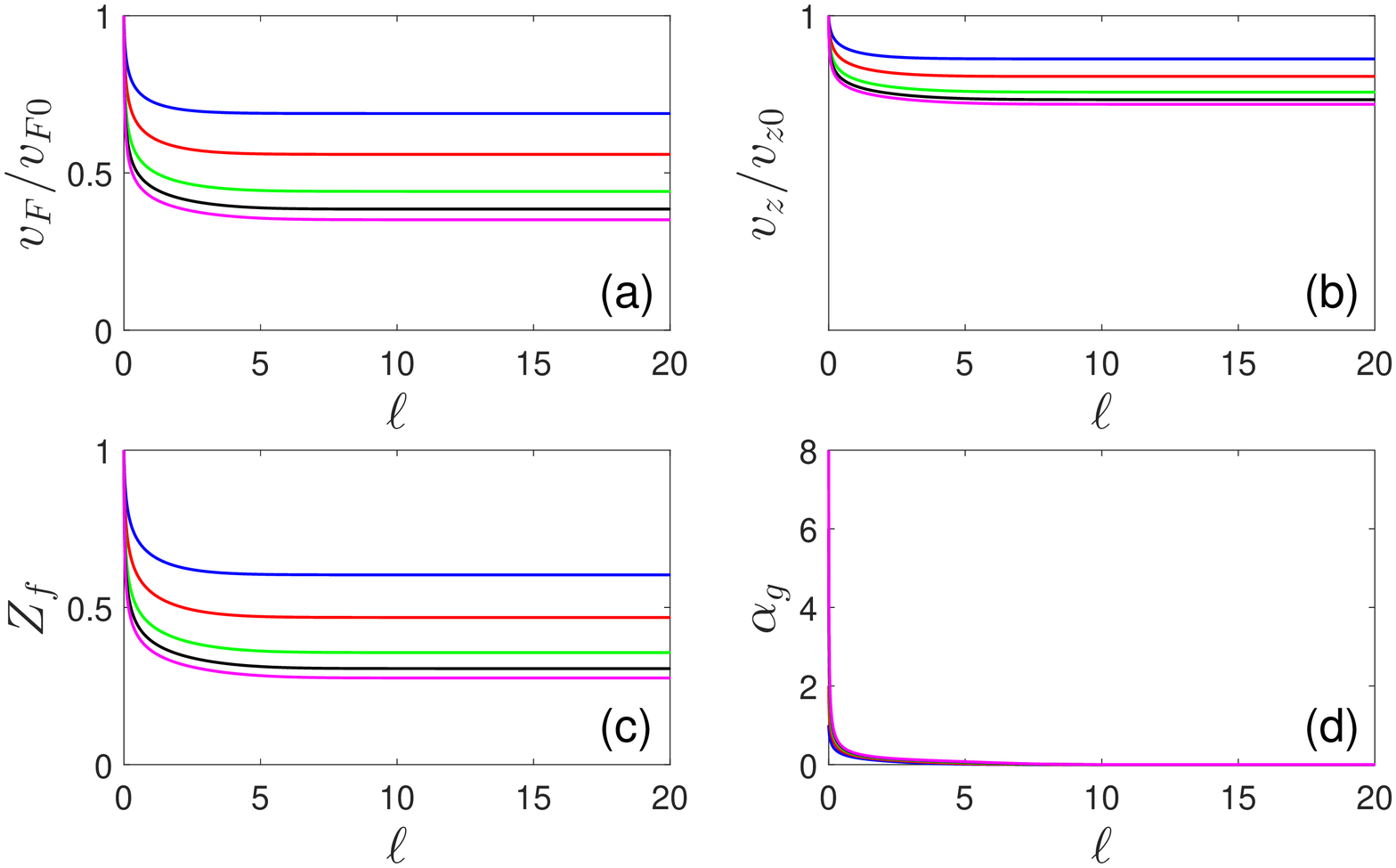}
\caption{(a)-(d) Flows of $v_{F}$, $v_{z}$, $Z_{f}$, $\alpha_{g}$.
Blue, red, green, black, magenta curves represent the initial values
$\alpha_{g0}=1, 2, 4, 6, 8$. The initial conditions
$\delta_{10}=0.1$, $\delta_{20}=0.05$, $\delta_{30}=0.05$, and
$\beta_{g0}=0.1$ are used in Figs.~\ref{Fig:VRGFermion} and
\ref{Fig:VRGBoson}. \label{Fig:VRGFermion}}
\end{figure}

\begin{figure}[htbp]
\center
\includegraphics[width=3.3in]{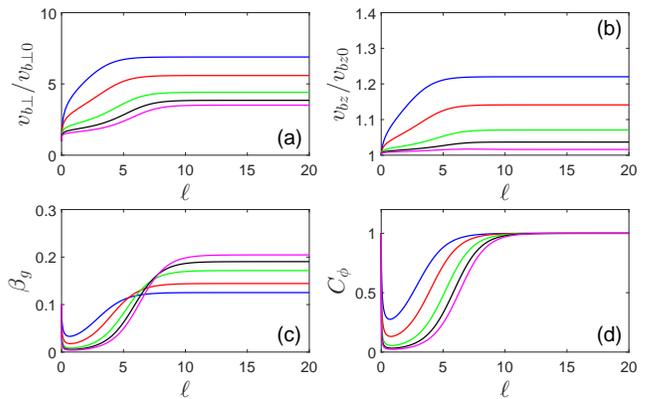}
\caption{(a)-(d) Flows of $v_{b\bot}$, $v_{bz}$, $\beta_{g}$, and
$C_{\phi}$, which characterize the impact of Yukawa coupling on the
excitonic fluctuation. Blue, red, green, black, magenta curves
represent the initial values $\alpha_{g0}=1, 2, 4, 6, 8$.
\label{Fig:VRGBoson}}
\end{figure}

The energy dependence of model parameters can be obtained by
numerically solving the above RG equations. Results are presented in
Figs.~\ref{Fig:VRGFermion} and \ref{Fig:VRGBoson}. For small initial
values $\delta_{10}$ and $\delta_{20}$, meaning that the motion of
boson (excitonic fluctuation) is slower than fermions, we see from
Fig.~\ref{Fig:VRGFermion}(a) that the fermion velocity $v_{F}$
decreases with lowering energy (growing $\ell$), and is saturated to
certain finite value $v_{F}^{*}$ eventually. According to
Fig.~\ref{Fig:VRGFermion}(a), the ratio $v_{F}^{*}/v_{F0}$ can be as
small as $0.4$ under certain circumstances. Since $v_F = k_F/m$, the
ratio between renormalized mass $m^{*}$ and bare mass $m$ is
inversely proportional to $v_{F}^{*}/v_{F0}$. For $\alpha_{g0} = 4$,
$m^{*}/m \approx 2.2$. As a comparison, the quantum oscillation
measurements of Pezzini \emph{et al.} \cite{Pezzini18} revealed that
$m^{*}/m$ is about $2$ in the limit of zero magnetic field. From
Fig.~\ref{Fig:VRGFermion}(b), we observe that $v_{z}$ is also
decreased by excitonic fluctuation. Therefore, our RG results
clearly show that slow excitonic quantum critical fluctuation could
lead to the unconventional fermion mass enhancement observed in
ZrSiS \cite{Pezzini18}.

As a system is approaching to a QCP, the renormalized quasiparticle
mass $m^{*}$ may be dramatically enhanced or even diverge
\cite{Rosch07, Ramshaw15, Taillefer19, Gegenwart02, Tokiwa13,
Aynajian12, Walmsley13}. In ZrSiS, the ratio $m^{*}/m \approx 2$.
While this turns out be the largest value ever observed in SM
materials \cite{Pezzini18}, it is much smaller than that observed in
the aforementioned unconventional superconductors \cite{Ramshaw15,
Taillefer19, Gegenwart02, Tokiwa13, Aynajian12, Walmsley13}. We now
explain why $m^{*}/m$ only takes a moderately large value. From
quantum many-body theory, we know that the ratio $m^{*}/m$ embodies
the importance of inter-particle interactions. In particular, it is
linked to the quasiparticle residue $Z_f$: $Z_f \sim m/m^{*}$. For a
system tuned to a QCP that exhibits non-Fermi liquid behavior, $Z_f
\rightarrow 0$ and $m^{*}$ diverges. For ZrSiS, we see from
Fig.~\ref{Fig:VRGFermion}(c) that the $\ell$-dependence of $Z_{f}$
is strongly reduced, implying the importance of excitonic
fluctuation. However, as $\ell$ goes to infinity, $Z_f$ does not
vanish but flows to a small finite value. On one hand, this result
indicates that ZrSiS is still a Fermi liquid, albeit a strongly
interacting one. On the other hand, it guarantees that the ratio
$m^{*}/m$ does not take a very large value in ZrSiS. In contrast,
the normal states of cuprate, heavy fermion, and iron-based
superconductors \cite{Rosch07, Ramshaw15, Taillefer19, Gegenwart02,
Tokiwa13, Aynajian12, Walmsley13} are known to be non-Fermi liquids,
thus it is not surprising that their mass enhancement is apparently
more significant than ZrSiS.

In the lowest energy limit, the anomalous dimension of fermion field
$\eta_{\psi} = C_{0}$ vanishes, and the effective strength of Yukawa
coupling $\alpha_{g}$ also flows to zero. Therefore, the excitonic
quantum fluctuation is an irrelevant perturbation at low energies.
However, before $\alpha_{g}$ goes to zero, the excitonic fluctuation
renormalizes the quasiparticle mass, leading to considerable mass
enhancement \cite{Pezzini18}.

The low-energy dynamics of the boson is also affected by the Yukawa
coupling. According to Figs.~\ref{Fig:VRGBoson}(a) and
\ref{Fig:VRGBoson}(b), $v_{b\bot}$ flows from initial value
$v_{b\bot0}$ to a larger constant value $v_{b\bot}^{*}$, and
$v_{bz}$ flows from $v_{bz0}$ to a larger constant $v_{bz}^{*}$.
Fig.~\ref{Fig:VRGBoson}(c) shows that $\beta_{g}$ approaches to a
finite value in the lowest energy limit. Accordingly, $C_{\phi}$,
$C_{\bot}$, and $C_{z}$ all flow to finite values. As shown in
Fig.~\ref{Fig:VRGBoson}(d), $C_{\phi}\rightarrow1$ in the limit
$\ell\rightarrow\infty$. Both $C_{\bot}$ and $C_{z}$ flow to unity
in the lowest energy limit. The bosonic anomalous dimension
$\eta_{\phi}$ is determined by $\eta_{\phi}=C_{\phi}^{*}=1$.

\begin{figure}[htbp]
\center
\includegraphics[width=3.3in]{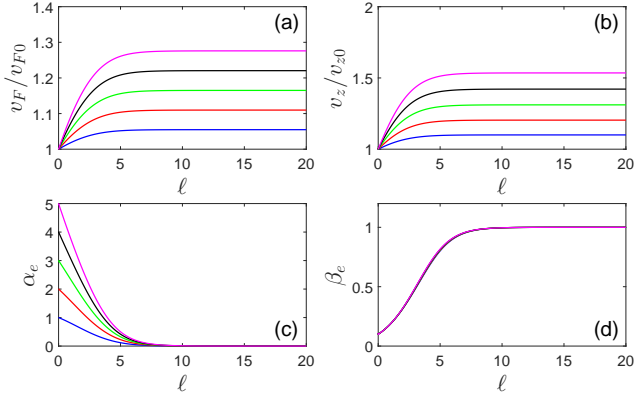}
\caption{(a)-(d) Flows of $v_{F}$, $v_{z}$, $\alpha_{e}$, and
$\beta_{e}$ caused purely by long-range Coulomb interaction. Blue,
red, green, black, magenta curves correspond to initial values
$\alpha_{e0}=1, 2, 3, 4, 5$. We choose $\beta_{e0}=0.1$, $a_{0}$=1,
and $\delta_{10}=0.2$. Apparently, Coulomb interaction increases
velocity $v_F$ and decreases mass. \label{Fig:VRGCoulomb}}
\end{figure}

\emph{Roles of LRCI and short-range interactions.} There are two
types of interactions in NLSMs: LRCI and short-range four-fermion
interaction. It is well established that the fermion velocity
renormalization unveiled in graphene  is produced by the LRCI
\cite{Elias11}. This might motivate one to speculate that the mass
enhancement of ZrSiS is also due to LRCI \cite{Pezzini18}. The
influence of LRCI on NLSM was studied by Huh \emph{et al.}
\cite{Huh16} and Wang and Nandkishore \cite{Wang17}, who found that
LRCI is irrelevant at low energies. However, whether LRCI increases
or reduces velocities was not explicitly answered in \cite{Huh16}
and \cite{Wang17}.

To determine the role played by LRCI, we performed RG analysis of
the dynamics of nodal line fermions subjected only to LRCI, with
details given in Supplementary Materials. The low-energy behaviors
of $v_{F}$, $v_{z}$, $\alpha_{e}$, and $\beta_{e}$ are presented in
Fig.~\ref{Fig:VRGCoulomb}, where $\alpha_{e}=e^{2}/v_{F}$ is the
effective strength of LRCI and
$\beta_{e}=\frac{\sqrt{2}e^{2}k_{F}}{32\pi \Lambda}$ embodies the
dynamical screening. The fermion velocities are always increased by
the LRCI. Thus LRCI tends to reduce, rather than enhance, the
quasiparticle mass of NLSMs. As shown in
Fig.~\ref{Fig:VRGCoulomb}(c), the strength parameter $\alpha_e $
flows to zero quickly with growing $\ell$, which is consistent with
previous work \cite{Huh16, Wang17}. Since the quasiparticle mass is
observed to be clearly enhanced in ZrSiS, we conclude that LRCI
plays little role and can be well ignored.

The short-range interactions are important and should be seriously
considered \cite{Roy17, Rudenko18}. We first emphasize that
short-range interaction cannot directly enhance fermion mass as they
do not renormalize fermion velocities \cite{Roy17}. However, it can
affect fermion mass in an indirect way: strong short-range
interaction leads to excitonic pairs of nodal line fermions; this
then drives the system to get sufficiently close to SM-EI QCP. Near
this QCP, slow exitonic quantum fluctuation enhances the fermion
mass.

\emph{Observable Quantities.} In order to distinguish the QCR from
the SM and EI phases, we now calculate a number of observable
quantities, including DOS, specific heat, compressibility, and
optical conductivities.

\begin{table*}[htbp]
\caption{Energy or temperature dependence of observable quantities
in the SM phase, the EI, and the QCR.}
\label{Table:ObservableQuantities} \vspace{-0.3cm}
\begin{center}
\setlength{\tabcolsep}{2mm}{
\begin{tabular}{|c|c|c|c|c|c|}
\hline\hline  & $\rho(\omega)$ & $C_{v}(T)$ & $\kappa(T)$ &
$\sigma_{\bot\bot}(|\Omega|)$ & $\sigma_{zz}(\Omega)$
\\
\hline SM & $|\omega|/(v_{F}v_{z})$ & $T^{2}/(v_{F}v_{z})$ &
$T/(v_{F}v_{z})$ & $v_{F}/v_{z}$ & $v_{z}/v_{F}$
\\
\hline QCR & $|\omega|/(v_{F}^{*}v_{z}^{*})$ &
$T^{2}/(v_{F}^{*}v_{z}^{*})$ & $T/(v_{F}^{*}v_{z}^{*})$
&$v_{F}^{*}/v_{z}^{*}$ & $v_{z}^{*}/v_{F}^{*}$
\\
\hline EI &
$|\omega|/(v_{F}v_{z})\theta\left(|\omega|-\Delta\right)$ &
$\frac{1}{v_{F}v_{z}}\frac{\Delta^{4}}{T^{2}}e^{-\frac{\Delta}{T}}$
& $\frac{1}{v_{F}v_{z}}\frac{\Delta^{2}}{T}e^{-\frac{\Delta}{T}}$ &
$\frac{v_{F}}{v_{z}} \left(1+\frac{4\Delta^{2}}{|\Omega|^{2}}\right)
\theta\left(|\Omega|-2\Delta\right)$ &
$\frac{v_{z}}{v_{F}}\left(1+\frac{4\Delta^{2}}{|\Omega|^{2}}\right)
\theta\left(|\Omega|-2\Delta\right)$
\\
\hline\hline
\end{tabular}}
\end{center}
\end{table*}

Deep in the NLSM phase, the fermion DOS is
\begin{eqnarray}
\rho(\omega) = \frac{Nk_{F}|\omega|}{2\pi v_{F}v_{z}},
\end{eqnarray}
and the specific heat depends on temperature as
\begin{eqnarray}
C_{v}(T) = \frac{9\zeta(3)Nk_{F}}{\pi v_{F}v_{z}}T^{2}.
\end{eqnarray}
The compressibility exhibits linear-in-$T$ behavior:
\begin{eqnarray}
\kappa(T) = \frac{2\ln(2)Nk_{F}}{\pi v_{F}v_{z}}T.
\end{eqnarray}
The optical conductivities within the $x$-$y$ plane and along the
$z$ axis are constants at low energies, namely
\begin{eqnarray}
\sigma_{\bot\bot}(\Omega) = \frac{Ne^{2}k_{F}}{32}\frac{v_{F}}{v_z},
\quad \sigma_{zz}(\Omega) = \frac{Ne^{2}k_{F}}{16}\frac{v_{z}}{v_F}.
\end{eqnarray}

At the SM-EI QCP, the velocities $v_{F}$ and $v_{z}$ are
renormalized to smaller values $v_{F}^{*}$ and $v_{z}^{*}$ by the
excitonic fluctuation, which then enhances DOS, specific heat, and
compressibility. Our RG results indicate that the ratio
$v_{F}^{*}/v_{z}^{*}$ is smaller than its bare value. Therefore, the
optical conductivity is suppressed within $x$-$y$ plane and is
enhanced along $z$ axis near the QCP.

If a finite excitonic gap $\Delta$ was generated on the Fermi
surface, all the above quantities would be significantly suppressed
at low energies. Actually, the DOS becomes
\begin{eqnarray}
\rho(\omega) = \frac{Nk_{F}|\omega|}{2\pi v_{F}v_{z}}
\theta\left(|\omega|-\Delta\right),
\end{eqnarray}
which is nonzero only above the energy scale set by $\Delta$. The
specific heat and compressibility are exponentially reduced by
$\Delta$ as follows
\begin{eqnarray}
C_{v}(T) &\approx& \frac{Nk_{F}}{\pi v_{F}v_{z}}
\frac{\Delta^4}{T^2}e^{-\frac{\Delta}{T}},
\\
\kappa(T) &\approx& \frac{Nk_{F}}{\pi v_{F}v_{z}}
\frac{\Delta^{2}}{T}e^{-\frac{\Delta}{T}}.
\end{eqnarray}
A discrete chiral symmetry is broken in EI phase, thus there is no
Goldstone boson. The specific heat is purely contributed by the
fermionic quasiparticles. The optical conductivities vanish at
energies below $2\Delta$:
\begin{eqnarray}
\sigma_{\bot\bot}(\Omega)&=&\frac{Ne^{2}k_{F}}{32}\frac{v_{F}}{v_{z}}
\left(1+\frac{4\Delta^{2}}{|\Omega|^{2}}\right)
\theta\left(|\Omega|-2\Delta\right),
\\
\sigma_{zz}(\Omega)&=&\frac{Ne^{2}k_{F}}{16}\frac{v_{z}}{v_{F}}
\left(1+\frac{4\Delta^{2}}{|\Omega|^{2}}\right)
\theta\left(|\Omega|-2\Delta\right).
\end{eqnarray}

In Table \ref{Table:ObservableQuantities}, we list the $\omega$- and
$T$-dependence of observable quantities obtained in the three
different regions of phase diagram Fig.~\ref{Fig:PhaseDiagram}. By
measuring these quantities, one could further verify whether ZrSiS
is near or far from the SM-EI QCP.

\emph{Summary and Discussion.} To summarize, we have studied the
behavior of gapless nodal line fermions near the QCP to an EI, and
found that the effective fermion mass is considerably enhanced by
the excitonic fluctuation. The quasiparticle residue $Z_f$ is
substantially suppressed, but flows to a finite value in the lowest
energy limit. Thus system should be identified as a strongly
correlated Fermi liquid. These results provide a clear explanation
of the unconventional mass enhancement in NLSM ZrSiS
\cite{Pezzini18}, and indicate that ZrSiS is a rare example of
topological SM with intriguing quantum criticality. Similar mass
enhancement might also occur in other NLSMs so long as the
short-range interaction is capable of driving the system
sufficiently close to a EI QCP.

The above theoretical analysis is carried out at SM-EI QCP with
$r=0$ and at zero chemical potential $\mu=0$. In real samples, $r$
and $\mu$ are usually not exactly zero. Here we comment on the
impact of finite $r$ and finite $\mu$. Finite $r$ weakens the
excitonic fluctuation, and thus leads to weaker enhancement of
quasiparticle mass. But the enhancement persists for small $r$. For
small $\mu$, theoretical results obtained at $\mu = 0$ are modified
only at energy scales below $\mu$. The excitoninc fluctuation can
still induce striking quantum critical behaviors, including strong
mass enhancement, in the QCR. On the experimental side, previous
ARPES measurements on ZrSiS indicated that the Dirac line nodes
connect the Dirac points near the Fermi level \cite{Neupane16,
FuBB19}, implying that $\mu$ should be quite small. Therefore, we
believe that our conclusion is qualitatively reliable for ZrSiS
\cite{Pezzini18}.

We acknowledge the support from the National Key R\&D Program of
China under Grants 2017YFA0403600 and 2016YFA0300404, and that from
the National Natural Science Foundation of China under Grants
11574285, 11504379, 11674327, 11974356, U1532267, and U1832209.

\onecolumngrid
\clearpage

\begin{center}
\textbf{\large{Supplementary Materials for ``\emph{Quantum Criticality of Excitonic Insulating Transition in Nodal Line Semimetal ZrSiS}"}}
\end{center}

In Section~\ref{Sec:DerivationYukawa}, we consider the
Yukawa-coupling between nodal line fermions and excitonic
fluctuation, and present the detailed derivation of the RG equations
of various model parameters. In Section~\ref{Sec:DerivationCoulomb},
we consider the long-range Coulomb interaction and derive the RG
equations of the corresponding model parameters.

\section{Influence of quantum fluctuation of excitonic insulating order
parameter \label{Sec:DerivationYukawa}}

\subsection{Propagators}

The propagator of free nodal line fermions in the imaginary
frequency formalism is given by
\begin{eqnarray}
G_{0}(\omega,\mathbf{k}) = \frac{1}{-i\omega + A\left(k_{\bot}^{2} -
k_{F}^{2}\right)\sigma_{1} + v_{z}k_{z}\sigma_{2}},
\label{Eq:FermionPropagatorOriginal}
\end{eqnarray}
where $A=1/(2m)$. In the low-energy regime, the fermion propagator
can be written as
\begin{eqnarray}
G_{0}(\omega,\mathbf{k}) = \frac{1}{-i\omega+v_{F}k_{r}\sigma_{1}
+v_{z}k_{z}\sigma_{2}}, \label{Eq:FermionPropagator}
\end{eqnarray}
where $v_{F}=k_{F}/m$ and $k_{r}=k_{\bot}-k_{F}$, and $\theta$ is
the angle between $\mathbf{k}_{\bot}$ and $\mathbf{q}_{\bot}$. The
bare propagator of boson, which describes the quantum fluctuation of
excitonic order parameter, takes the form
\begin{eqnarray}
D_{0}(\Omega,\mathbf{q})=\frac{1}{\Omega^{2}+v_{b\bot}^{2}q_{\bot}^{2}
+ v_{bz}^{2}q_{z}^{2}}. \label{Eq:BosonPropagator}
\end{eqnarray}

\subsection{Self-energy of the boson}

The self-energy of boson is defined as
\begin{eqnarray}
\Pi(\Omega,\mathbf{q})=-Ng^{2}\int\frac{d\omega}{2\pi}
\int'\frac{d^3\mathbf{k}}{(2\pi)^{3}}
\mathrm{Tr}\left[\sigma_{3}G_{0}(\omega,\mathbf{k})\sigma_{3}
G_{0}\left(\omega+\Omega,\mathbf{k}+\mathbf{q}\right)\right].
\label{Eq:BosonSelfEnergyDefinition}
\end{eqnarray}
Substituting Eq.~(\ref{Eq:FermionPropagatorOriginal}) into
Eq.~(\ref{Eq:BosonSelfEnergyDefinition}), we get
\begin{eqnarray}
\Pi(\Omega,\mathbf{q}) &=& 2Ng^{2}\int\frac{d\omega}{2\pi}
\int'\frac{d^3\mathbf{k}}{(2\pi)^{3}} \frac{\omega(\omega+\Omega) +
A^{2}2k_{F}k_{r}\left(2k_{F}k_{r}+2k_{F}q_{\bot}\cos(\theta) +
q_{\bot}^{2}\right)+v_{z}^{2}k_{z}(k_{z}+q_{z})}{\left(\omega^{2} +
E_{\mathbf{k}}^{2}\right)\left[(\omega+\Omega)^{2} + A^{2}
\left(2k_{F}k_{r}+2k_{F}q_{\bot}\cos(\theta)+q_{\bot}^{2}\right)^{2}
+ v_{z}^{2}(k_{z}+q_{z})^{2}\right]},
\end{eqnarray}
where $E_{\mathbf{k}}=\sqrt{v_{F}^{2}k_{r}^{2}+v_{z}^{2}k_{z}^{2}}$.
The prime in $\int'$ means that a proper momentum shell will be
chosen in the calculations. Expanding $\Pi(\Omega,\mathbf{q})$ to
the quadratic order of $\Omega$, $q_{r}$, and $q_{z}$, and then
performing the integration of $\omega$, we find
\begin{eqnarray}
\Pi(\Omega,\mathbf{q}) &=& -\Omega^{2}N\frac{g^{2}}{4} \int'
\frac{d^3\mathbf{k}}{(2\pi)^{3}}\frac{1}{E^{3}(\mathbf{k})} -
v_{r}^{2}q_{r}^{2}N\frac{g^{2}}{2}\int'\frac{d^3\mathbf{k}}{(2\pi)^{3}}
\left[\frac{\cos^{2}(\theta)}{E^{3}(\mathbf{k})} -
\frac{3v_{r}^{2}k_{r}^{2} \cos^{2}(\theta)}{2
E^{5}(\mathbf{k})}\right] \nonumber \\
&&-v_{z}^{2}q_{z}^{2}N\frac{g^{2}}{2}\int'\frac{d^3\mathbf{k}}{(2\pi)^{3}}
\left[\frac{1}{E^{3}(\mathbf{k})}-\frac{3v_{z}^{2}
k_{z}^{2}}{2E^{5}(\mathbf{k})}\right].
\end{eqnarray}
A redefinition $\Pi(\Omega,\mathbf{q})-\Pi(0,0) \rightarrow
\Pi(\Omega,\mathbf{q})$ has been employed to discard a constant
term, such that $\Pi(0,0)=0$ is fulfilled. Adopting the relation
\begin{eqnarray}
\int'\frac{d^3\mathbf{k}}{(2\pi)^{3}}\approx
k_{F}\int'\frac{dk_{r}dk_{z}}{(2\pi)^{2}},
\end{eqnarray}
and choosing the following momentum shell
\begin{eqnarray}
b\Lambda < \sqrt{v_{F}^{2}k_{r}^{2}+v_{z}^{2}k_{z}^{2}} < \Lambda,
\label{Eq:RGSchemeFermion}
\end{eqnarray}
where $b=e^{-\ell}$ with $\ell$ being RG running parameter, we
obtain
\begin{eqnarray}
\Pi(\Omega,\mathbf{q}) = -\Omega^{2}C_{\phi}\ell - v_{b\bot}^{2}
q_{\bot}^{2}C_{\bot}\ell-v_{bz}^{2}q_{z}^{2} C_{z}\ell,
\end{eqnarray}
where
\begin{eqnarray}
C_{\phi}=\frac{Ng^{2}k_{F}}{8\pi v_{r}v_{z}\Lambda},\quad
C_{\bot}=\frac{Ng^{2}k_{F}v_{r}}{32\pi v_{z}v_{b\bot}^{2}\Lambda},\quad
C_{z}=\frac{Ng^{2}k_{F}v_{z}}{16\pi v_{r}v_{bz}^{2}\Lambda}.
\end{eqnarray}

\subsection{Self-energy of the nodal line fermion}

The self-energy of nodal line fermions induced by excitonic
fluctuation is
\begin{eqnarray}
\Sigma(\omega,k_{F}+\mathbf{k}) = g^{2}\int'\frac{d\Omega}{2\pi}
\frac{d^3\mathbf{q}}{(2\pi)^{3}} \sigma_{3}
G_{0}(\omega+\Omega,\mathbf{k}+\mathbf{q})\sigma_{3}
D_{0}(\Omega,\mathbf{q}). \label{Eq:SelfEnergyFermionDef}
\end{eqnarray}
Here, the momentum vector $\mathbf{k} = (k_{x},0,k_{z})$.
Substituting Eq.~(\ref{Eq:FermionPropagator}) into
Eq.~(\ref{Eq:SelfEnergyFermionDef}) gives rise to
\begin{eqnarray}
\Sigma(\omega,k_{F}+\mathbf{k}) = g^{2}\int'\frac{d\Omega}{2\pi}
\frac{d^3\mathbf{q}}{(2\pi)^{3}}\frac{i\left(\omega+\Omega\right) -
v_{F}\left(k_{x}+q_{x}\right)\sigma_{1} -
v_{z}\left(k_{z}+q_{z}\right)\sigma_{2}}{\left(\omega+\Omega\right)^{2}
+ v_{F}^{2}\left(k_{x}+q_{x}\right)^{2} + v_{z}^{2}
\left(k_{z}+q_{z}\right)^{2}}D_{0}(\Omega,\mathbf{q}).
\end{eqnarray}
We then expand this function to the leading orders of $\Omega$ and
$\mathbf{k}_{i}$, and insert the expression of
$D_{0}(\Omega,\mathbf{q})$, which yields
\begin{eqnarray}
\Sigma(\omega,k_{F}+\mathbf{k}) &=& i\omega\frac{g^{2}}{16\pi^{4}}
\int' d\Omega dq_{x}dq_{y}dq_{z} \frac{-\Omega^{2} + v_{F}^{2}
q_{x}^{2} + v_{z}^{2}q_{z}^{2}}{\left(\Omega^{2}+v_{F}^{2}q_{x}^{2}
+ v_{z}^{2}q_{z}^{2}\right)^{2}} \frac{1}{\Omega^{2}+v_{b\bot}^{2}
\left(q_{x}^{2}+q_{y}^{2}\right)+v_{bz}^{2}q_{z}^{2}}\nonumber \\
&&-v_{F}k_{x}\sigma_{1}\frac{g^{2}}{16\pi^{4}}\int'd\Omega dq_{x}
dq_{y}dq_{z}\frac{\Omega^{2}-v_{F}^{2}q_{x}^{2}+v_{z}^{2}
q_{z}^{2}}{\left(\Omega^{2}+v_{F}^{2}q_{x}^{2} +
v_{z}^{2}q_{z}^{2}\right)^{2}}\frac{1}{\Omega^{2}+v_{b\bot}^{2}
\left(q_{x}^{2}+q_{y}^{2}\right) +v_{bz}^{2}q_{z}^{2}}\nonumber
\\
&&-v_{z}k_{z}\sigma_{2}\frac{g^{2}}{16\pi^{4}}\int'd\Omega
dq_{x}dq_{y}dq_{z}\frac{\Omega^{2}+v_{F}^{2}q_{x}^{2}-v_{z}^{2}
q_{z}^{2}}{\left(\Omega^{2}+v_{F}^{2}q_{x}^{2}+v_{z}^{2}
q_{z}^{2}\right)^{2}} \frac{1}{\Omega^{2}+v_{b\bot}^{2}
\left(q_{x}^{2}+q_{y}^{2}\right)+v_{bz}^{2}q_{z}^{2}}.
\end{eqnarray}
Perform integration as follows
\begin{eqnarray}
\int'd\Omega dq_{x}dq_{y}dq_{z}=\left(\int_{b\Lambda}^{\Lambda} +
\int_{-\Lambda}^{-b\Lambda}\right)dq_{x}\int_{-\infty}^{+\infty}
d\Omega \int_{-\infty}^{+\infty}dq_{y}
\int_{-\infty}^{+\infty}dq_{z}, \label{Eq:RGSchemeBoson}
\end{eqnarray}
we obtain
\begin{eqnarray}
\Sigma(\omega,k_{F}+\mathbf{k}) = i\omega C_{0}\ell-v_{F}k_{x}
\sigma_{1}C_{1}\ell-v_{z}k_{z}\sigma_{2}C_{2}\ell,
\end{eqnarray}
where the constants $C_{1,2,3}$ are
\begin{eqnarray}
C_{0}&=&\frac{g^{2}}{8\pi^{3}v_{F}^{3}\delta_{2}^{2}}
\int_{-\infty}^{+\infty}dx\int_{-\infty}^{+\infty}dy
\frac{-x^{2}+1+\delta_{1}^{2}y^{2}}{\left(x^{2}+1+\delta_{1}^{2}
y^{2}\right)^{2}} \frac{1}{\sqrt{\frac{1}{\delta_{2}^{2}}x^{2} + 1 +
\frac{\delta_{3}^{2}}{\delta_{2}^{2}}y^{2}}}, \\
C_{1}&=&\frac{g^{2}}{8\pi^{3}v_{F}^{3}\delta_{2}^{2}}
\int_{-\infty}^{+\infty}dx\int_{-\infty}^{+\infty}dy
\frac{x^{2}-1+\delta_{1}^{2}y^{2}}{\left(x^{2}+1+\delta_{1}^{2}
y^{2}\right)^{2}} \frac{1}{\sqrt{\frac{1}{\delta_{2}^{2}}x^{2}+1
+ \frac{\delta_{3}^{2}}{\delta_{2}^{2}}y^{2}}}, \\
C_{2}&=&\frac{g^{2}}{8\pi^{3}v_{F}^{3}\delta_{2}^{2}}
\int_{-\infty}^{+\infty}dx\int_{-\infty}^{+\infty}dy
\frac{x^{2}+1-\delta_{1}^{2}y^{2}}{\left(x^{2}+1+\delta_{1}^{2}
y^{2}\right)^{2}} \frac{1}{\sqrt{\frac{1}{\delta_{2}^{2}}x^{2}+1 +
\frac{\delta_{3}^{2}}{\delta_{2}^{2}}y^{2}}},
\end{eqnarray}
with
\begin{eqnarray}
\delta_{1}=\frac{v_{z}}{v_{F}},\qquad
\delta_{2}=\frac{v_{b\bot}}{v_{F}},\qquad
\delta_{3}=\frac{v_{bz}}{v_{F}}.
\end{eqnarray}

\subsection{Vertex correction to fermion-boson (Yukawa) coupling}

The correction to the fermion-boson coupling is
\begin{eqnarray}
\Gamma = g^{2}\int'\frac{d\Omega}{2\pi}\frac{d^3
\mathbf{q}}{(2\pi)^{3}}\sigma_{3}G_{0}(\Omega,\mathbf{q})\sigma_{3}
G_{0}(\Omega,\mathbf{q})\sigma_{3} D_{0}(\Omega,\mathbf{q}).
\label{Eq:VortexCorrectionYukawaDefinition}
\end{eqnarray}
Substituting Eqs.~(\ref{Eq:FermionPropagator}) and
(\ref{Eq:BosonPropagator}) into
Eq.~(\ref{Eq:VortexCorrectionYukawaDefinition}), we obtain
\begin{eqnarray}
\Gamma = -\sigma_{3}C_{3}\ell,
\end{eqnarray}
where
\begin{eqnarray}
C_{3} = \frac{g^{2}}{8\pi^{3}v_{F}^{3}\delta_{2}^{2}}
\int_{-\infty}^{+\infty}dx\int_{-\infty}^{+\infty}dy
\frac{1}{x^{2}+1 +\delta_{1}^{2}y^{2}}
\frac{1}{\sqrt{\frac{1}{\delta_{2}^{2}}x^{2}
+1+\frac{\delta_{3}^{2}}{\delta_{2}^{2}}y^{2}}}.
\end{eqnarray}

\subsection{Correction to $\phi^{4}$ coupling}

The correction to the $\phi^{4}$ vertex induced by $\phi^{4}$
coupling is
\begin{eqnarray}
\delta u^{a} = -\frac{3}{2}u^{2}\int'\frac{d\Omega}{2\pi}
\frac{d^3\mathbf{q}}{(2\pi)^{3}}D_{0}(\Omega,\mathbf{q})
D_{0}(\Omega,\mathbf{q}).\label{Eq:Phi4CorrectionADefinition}
\end{eqnarray}
Making use of Eq.~(\ref{Eq:BosonPropagator}) into
Eq.~(\ref{Eq:Phi4CorrectionADefinition}), we now re-write $\delta
u^{a}$ as
\begin{eqnarray}
\delta u^{a} = -\frac{3u^{2}}{32\pi^{4}}\int'd\Omega
dq_{x}dq_{y}dq_{z}\frac{1}{\left(\Omega^{2}+v_{b\bot}^{2}
q_{\bot}^{2}+v_{bz}^{2}q_{z}^{2}\right)^{2}},
\end{eqnarray}
which after carrying out integration becomes
\begin{eqnarray}
\delta u^{a} = -\frac{3u^{2}}{16\pi^{2}v_{b\bot}^{3}}
\frac{\delta_{2}}{\delta_{3}}\ell.
\end{eqnarray}

The correction to $\phi^{4}$ vertex due to the fermion-boson
coupling can be computed as follows
\begin{eqnarray}
\delta u^{b} &=& 6g^{4}N\int'\frac{d^3\mathbf{k}}{(2\pi)^{3}}
\frac{d\omega}{2\pi} \mathrm{Tr}\left[\sigma_{3}
G_{0}(\omega,\mathbf{k})\sigma_{3}G_{0}(\omega,\mathbf{k})
\sigma_{3}G_{0}(\omega,\mathbf{k})\sigma_{3}
G_{0}(\omega,\mathbf{k})\right]\nonumber \label{Eq:Phi4CorrectionBDefinition}
\\
&=& N\frac{3g^{4}k_{F}}{4\pi}\int'dk_{r}dk_{z}
\frac{1}{\left(v_{F}^{2}k_{r}^{2}+v_{z}^{2}k_{z}^{2}
\right)^{\frac{3}{2}}} \nonumber\\
&=& N\frac{3g^{4}k_{F}}{2 v_{F}v_{z}\Lambda}\ell.
\end{eqnarray}

Now the total correction to $\phi^{4}$ vertex has the form
\begin{eqnarray}
\delta u = \delta u^{a}+\delta u^{b} = \left(-\frac{3u^{2}}{16
\pi^{2}v_{b\bot}^{3}} \frac{\delta_{2}}{\delta_{3}} +
N\frac{3g^{4}k_{F}}{2 v_{F}v_{z}\Lambda}\right)\ell.
\end{eqnarray}

\subsection{Derivation of the RG equations}

The free action of fermion field $\psi$ is
\begin{eqnarray}
S_{\psi}=\int\frac{d\omega}{2\pi}\frac{dk_{x}}{2\pi}\frac{dk_{y}}{2\pi}
\frac{dk_{z}}{2\pi}\psi_{a}^{\dag}(\omega,\mathbf{k}) \left(-i\omega
+ \frac{k_{x}^{2}+k_{y}^{2}-k_{F}^{2}}{2m}\sigma_{x} + v_{z}k_{z}
\sigma_{y}\right)\psi_{a}(\omega,\mathbf{k}).
\end{eqnarray}
In the low-energy regime, this free action can be re-written as
\begin{eqnarray}
S_{\psi}=k_{F}\int\frac{d\omega}{2\pi}\frac{dk_{r}}{2\pi}
\frac{dk_{z}}{2\pi}\psi_{a}^{\dag}(\omega,\mathbf{k})
\left(-i\omega+v_{F}k_{r}\sigma_{1}+v_{z}k_{z}\sigma_{2}
\right)\psi_{a}(\omega,\mathbf{k}).\label{Eq:ActionNodalFermions}
\end{eqnarray}
Adding the fermion self-energy to the above free action leads to
\begin{eqnarray}
S_{\psi} &=& k_{F}\int\frac{d\omega}{2\pi}\frac{dk_{r}}{2\pi}
\frac{dk_{z}}{2\pi}\psi_{a}^{\dag}(\omega,\mathbf{k})
\left(-i\omega+v_{F}k_{r}\sigma_{1} + v_{z}k_{z}\sigma_{2} -
\Sigma(\omega,\mathbf{k})\right)\psi_{a}(\omega,\mathbf{k})\nonumber
\\
&=&k_{F}\int\frac{d\omega}{2\pi}\frac{dk_{r}}{2\pi}
\frac{dk_{z}}{2\pi}\psi_{a}^{\dag}(\omega,\mathbf{k})\left(-i\omega
e^{C_{0}\ell}+v_{F}k_{r}\sigma_{1}e^{C_{1}\ell}+v_{z}k_{z}
\sigma_{2} e^{C_{2}\ell}\right)\psi_{a}(\omega,\mathbf{k}).
\end{eqnarray}
Make the following scaling transformations:
\begin{eqnarray}
\omega&=&\omega'e^{-\ell}, \label{Eq:ScalingomegaF}
\\
k_{r}&=&k_{r}'e^{-\ell}, \label{Eq:Scalingkr}
\\
k_{z}&=&k_{z}'e^{-\ell}, \label{Eq:Scalingkz}
\\
\psi&=&\psi' e^{\frac{\left(4-C_{0}\right)}{2}\ell},
\label{Eq:Scalingpsi}
\\
v_{F}&=&v_{F}'e^{-\left(C_{1}-C_{0}\right)\ell},
\label{Eq:ScalingvF}
\\
v_{z}&=&v_{z}'e^{-\left(C_{2}-C_{0}\right)\ell},
\label{Eq:Scalingvz}
\end{eqnarray}
and then express the total action in the form
\begin{eqnarray}
S_{\psi'} = k_{F}\int\frac{d\omega'}{2\pi}\frac{dk_{r}'}{2\pi}
\frac{dk_{z}'}{2\pi}\psi_{a}'^{\dag}(\omega',\mathbf{k}')
\left(-i\omega'+v_{F}'k_{r}'\sigma_{x} +v_{z}'k_{z}'\sigma_{y}
\right)\psi_{a}'(\omega',\mathbf{k}'),
\end{eqnarray}
which recovers the original form of the fermion action.

The free action of boson field $\phi$ is given by
\begin{eqnarray}
S_{\phi} = \int\frac{d\Omega}{2\pi}\frac{dq_{x}}{2\pi}
\frac{dq_{y}}{2\pi}\frac{dq_{z}}{2\pi}\phi(\Omega,\mathbf{q})
\left(\Omega^{2}+v_{b\bot}^{2}q_{\bot}^{2}+v_{bz}^{2}q_{z}^{2}
\right)\phi(\Omega,\mathbf{q}).
\end{eqnarray}
Including the boson self-energy modifies it into
\begin{eqnarray}
S_{\phi} &=& \int\frac{d\Omega}{2\pi}\frac{dq_{x}}{2\pi}
\frac{dq_{y}}{2\pi}\frac{dq_{z}}{2\pi}\phi(\Omega,\mathbf{q})
\left(\Omega^{2}+v_{b\bot}^{2}q_{\bot}^{2}+v_{bz}^{2}q_{z}^{2} -
\Pi(\Omega,\mathbf{q})\right)\phi(\Omega,\mathbf{q})\nonumber
\\
&\approx&\int\frac{d\Omega}{2\pi}\frac{dq_{x}}{2\pi}
\frac{dq_{y}}{2\pi}\frac{dq_{z}}{2\pi}\phi(\Omega,\mathbf{q})
\left(\Omega^{2}e^{C_{\phi}\ell}+v_{b\bot}^{2}q_{\bot}^{2}e^{C_{\bot}\ell}
+v_{bz}^{2}q_{z}^{2}e^{C_{z}\ell}\right)\phi(\Omega,\mathbf{q}).
\end{eqnarray}

To proceed, we make the following scaling transformations
\begin{eqnarray}
\Omega&=&\Omega'e^{-\ell}, \label{Eq:ScalingOmegaB}
\\
q_{x}&=&q_{x}'e^{-\ell}, \label{Eq:Scalingqx}
\\
q_{y}&=&q_{y}'e^{-\ell}, \label{Eq:Scalingqy}
\\
q_{z}&=&q_{z}'e^{-\ell}, \label{Eq:Scalingqz}
\\
\phi&=&\phi'e^{\frac{\left(6-C_{\phi}\right)}{2}\ell},
\label{Eq:Scalingphi}
\\
v_{b\bot}&=&v_{b\bot}'e^{-\frac{\left(C_{\bot}-C_{\phi}\right)}{2}\ell},
\label{Eq:Scalingvbbot}
\\
v_{bz}&=&v_{bz}'e^{-\frac{\left(C_{z}-C_{\phi}\right)}{2}\ell}.
\label{Eq:Scalingvbz}
\end{eqnarray}
The modified (by interaction) boson action now has the form
\begin{eqnarray}
S_{\phi'} = \int\frac{d\Omega'}{2\pi}\frac{dq_{x}'}{2\pi}
\frac{dq_{y}'}{2\pi}\frac{dq_{z}'}{2\pi}\phi'(\Omega',\mathbf{q}')
\left(\Omega'^{2}+v_{b\bot}'^{2}q_{\bot}'^{2} + v_{bz}'^{2}
q_{z}'^{2}\right)\phi'(\Omega',\mathbf{q}'),
\end{eqnarray}
which has same form as the original action of boson.

The action of the fermion-boson coupling is originally defined as
\begin{eqnarray}
S_{\psi\phi} = g\int\frac{d\omega}{2\pi}\frac{d^3\mathbf{k}}{(2
\pi)^{3}} \frac{d\Omega}{2\pi} \frac{d^3\mathbf{q}}{(2\pi)^{3}}
\psi_{a}^{\dag}(\omega,\mathbf{k})\sigma_{3}
\psi_{a}(\omega+\Omega,\mathbf{k}+\mathbf{q})\phi(\Omega,\mathbf{q}),
\end{eqnarray}
which can be further written as
\begin{eqnarray}
S_{\psi\phi} = gk_{F}\int\frac{d\omega}{2\pi}\frac{dk_{r}}{2\pi}
\frac{dk_{z}}{2\pi}\frac{d\Omega}{2\pi}
\frac{d^3\mathbf{q}}{(2\pi)^{3}}\psi_{a}^{\dag}(\omega,\mathbf{k})
\sigma_{3}\psi_{a}(\omega+\Omega,\mathbf{k}+\mathbf{q})\phi(\Omega,\mathbf{q}).
\end{eqnarray}
This action is changed by the vertex correction to become
\begin{eqnarray}
S_{\psi\phi} &=& g\left(1-C_{3}\ell\right)k_{F}\int
\frac{d\omega}{2\pi} \frac{dk_{r}}{2\pi}\frac{dk_{z}}{2\pi}
\frac{d\Omega}{2\pi} \frac{d^3\mathbf{q}}{(2\pi)^{3}}
\psi_{a}^{\dag}(\omega,\mathbf{k})\sigma_{3}
\psi_{a}(\omega+\Omega,\mathbf{k}+\mathbf{q})\phi(\Omega,\mathbf{q})\nonumber
\\
&\approx& g e^{-C_{3}\ell}k_{F}\int\frac{d\omega}{2\pi}
\frac{dk_{r}}{2\pi} \frac{dk_{z}}{2\pi}\frac{d\Omega}{2\pi}
\frac{d^3\mathbf{q}}{(2\pi)^{3}}\psi_{a}^{\dag}(\omega,\mathbf{k})\sigma_{3}
\psi_{a}(\omega+\Omega,\mathbf{k}+\mathbf{q})\phi(\Omega,\mathbf{q}).
\end{eqnarray}
Performing the scaling transformations given by
Eqs.~(\ref{Eq:ScalingomegaF})-(\ref{Eq:Scalingpsi}), and
Eqs.~(\ref{Eq:ScalingOmegaB})-(\ref{Eq:Scalingphi}), we finally
expression the above action as
\begin{eqnarray}
S_{\psi'\phi'} &=& gk_{F}e^{-\left(C_{0}+C_{3} +
\frac{C_{\phi}}{2}\right)\ell} \int\frac{d\omega'}{2\pi}
\frac{dk_{r}'}{2\pi}\frac{dk_{z}'}{2\pi}\frac{d\Omega'}{2\pi}
\frac{d^3\mathbf{q}'}{(2\pi)^{3}}\psi_{a}'^{\dag}(\omega',\mathbf{k}')
\sigma_{3}\psi_{a}'(\omega'+\Omega',\mathbf{k}'+\mathbf{q}')
\phi'(\Omega',\mathbf{q}')\nonumber\\
&=& g'k_{F} \int\frac{d\omega'}{2\pi}\frac{dk_{r}'}{2\pi}
\frac{dk_{z}'}{2\pi}\frac{d\Omega'}{2\pi}
\frac{d^3\mathbf{q}'}{(2\pi)^{3}}\psi_{a}'^{\dag}(\omega',\mathbf{k}')
\sigma_{3} \psi_{a}'(\omega'+\Omega',\mathbf{k}'+\mathbf{q}')
\phi'(\Omega',\mathbf{q}'),
\end{eqnarray}
where the following transformation is made for coupling parameter
$g$:
\begin{eqnarray}
g = g'ke^{\left(C_{0}+C_{3}+\frac{C_{\phi}}{2}\right)\ell}.
\label{Eq:Scalingg}
\end{eqnarray}

The action of $\phi^4$ coupling is originally defined as
\begin{eqnarray}
S_{\phi^{4}}&=&\frac{u}{24}\int\frac{d\Omega_{1}}{2\pi}
\frac{d^3\mathbf{q}_{1}}{(2\pi)^{3}}
\frac{d\Omega_{2}}{2\pi}\frac{d^3\mathbf{q}_{2}}{(2\pi)^{3}}
\frac{d\Omega_{3}}{2\pi}\frac{d^3\mathbf{q}_{3}}{(2\pi)^{3}}
\phi(\Omega_{1},\mathbf{q}_{1})\phi(\Omega_{2},\mathbf{q}_{2})
\phi(\Omega_{3},\mathbf{q}_{3})\nonumber
\\
&&\times\phi(-\Omega_{1}-\Omega_{2}-\Omega_{3},
-\mathbf{q}_{1}-\mathbf{q}_{2}-\mathbf{q}_{3}).
\end{eqnarray}
The same calculational steps can be repeated to obtain
\begin{eqnarray}
S_{\phi'^{4}} &\approx& \frac{1}{24}\left[u-2C_{\phi}u\ell +
\left(-\frac{3u^{2}}{16\pi^{2}v_{b\bot}^{2}v_{bz}} + N
\frac{3g^{4}k_{F}}{2v_{F}v_{z}\Lambda}\right)\ell\right]\int\frac{d\Omega_{1}'}{2\pi}\frac{d^3\mathbf{q}_{1}'}{(2\pi)^{3}}
\frac{d\Omega_{2}'}{2\pi}\frac{d^3\mathbf{q}_{2}'}{(2\pi)^{3}}
\frac{d\Omega_{3}'}{2\pi}\frac{d^3\mathbf{q}_{3}'}{(2\pi)^{3}}
\phi'(\Omega_{1}',\mathbf{q}_{1}')\phi'(\Omega_{2}',\mathbf{q}_{2}')\nonumber
\\
&&\times\phi'(\Omega_{3}',\mathbf{q}_{3}')\phi'(-\Omega_{1}'-\Omega_{2}'
-\Omega_{3}',-\mathbf{q}_{1}'-\mathbf{q}_{2}'-\mathbf{q}_{3}').
\end{eqnarray}
Redefining the coupling parameter
\begin{eqnarray}
u' = u-2C_{\phi}u\ell+\left(-\frac{3u^{2}}{16\pi^{2}
v_{b\bot}^{2}v_{bz}} +N\frac{3g^{4}k_{F}}{2 v_{F}v_{z}
\Lambda}\right)\ell, \label{Eq:Scalingu}
\end{eqnarray}
we get
\begin{eqnarray}
S_{\phi'^{4}} &\approx& u' \int\frac{d\Omega_{1}'}{2\pi}
\frac{d^3\mathbf{q}_{1}'}{(2\pi)^{3}} \frac{d\Omega_{2}'}{2\pi}
\frac{d^3\mathbf{q}_{2}'}{(2\pi)^{3}}
\frac{d\Omega_{3}'}{2\pi}\frac{d^3\mathbf{q}_{3}'}{(2\pi)^{3}}
\phi'(\Omega_{1}',\mathbf{q}_{1}')\phi'(\Omega_{2}',\mathbf{q}_{2}')
\phi'(\Omega_{3}',\mathbf{q}_{3}')\nonumber
\\
&&\times\phi'(-\Omega_{1}'-\Omega_{2}'
-\Omega_{3}',-\mathbf{q}_{1}'-\mathbf{q}_{2}'-\mathbf{q}_{3}').
\end{eqnarray}

From the transformations of Eqs.~(\ref{Eq:ScalingvF}),
(\ref{Eq:Scalingvz}), (\ref{Eq:Scalingvbbot}),
(\ref{Eq:Scalingvbz}), (\ref{Eq:Scalingg}), (\ref{Eq:Scalingu}), we
eventually derive the self-consistently coupled RG equations
\begin{eqnarray}
\frac{dv_{F}}{d\ell}&=&\left(C_{1}-C_{0}\right)v_{F},
\\
\frac{dv_{z}}{d\ell}&=&\left(C_{2}-C_{0}\right)v_{z},
\\
\frac{dv_{b\bot}}{d\ell}&=&\frac{\left(C_{\bot}-C_{\phi}\right)}{2}v_{b\bot},
\\
\frac{dv_{bz}}{d\ell}&=&\frac{\left(C_{z}-C_{\phi}\right)}{2}v_{bz},
\\
\frac{d\delta_{1}}{d\ell} &=& \left(C_{2}-C_{1}\right)\delta_{1},
\\
\frac{d\delta_{2}}{d\ell}&=&\left(\frac{C_{\bot}-C_{\phi}}{2} -
C_{1}+C_{0}\right)\delta_{2},
\\
\frac{d\delta_{3}}{d\ell}&=&\left(\frac{C_{z}-C_{\phi}}{2} -
C_{1}+C_{0}\right)\delta_{3},
\\
\frac{d\alpha_{g}}{d\ell} &=&
-\left(-C_{0}+3C_{1}+2C_{3}+C_{\phi}\right)\alpha_{g},
\\
\frac{d\beta_{g}}{d\ell}&=&\left(1-C_{\phi}-2C_{1}-2C_{3}\right)\beta_{g},
\\
\frac{du}{d\ell} &=& -\left(C_{\bot}+\frac{C_{z}}{2} +
\frac{C_{\phi}}{2}\right)u - \frac{3u^{2}}{16} + \frac{12\alpha_{g}
\beta_{g}}{\pi \delta_{1}\delta_{2}^{2}\delta_{3}},
\end{eqnarray}
where
\begin{eqnarray}
\alpha_{g}=\frac{g^{2}}{v_{F}^{3}},\qquad
\beta_{g}=\frac{Ng^{2}k_{F}}{8\pi v_{F}^{2}\Lambda}.
\end{eqnarray}
In the derivation, we have made the redefinition
\begin{eqnarray}
\frac{u}{\pi^{2} v_{b\bot}^{2}v_{bz}}\rightarrow u.
\end{eqnarray}
The parameters $C_{0,1,2,3}$, $C_{\phi}$, $C_{\bot}$, and $C_{z}$
appearing in the RG equations are given by
\begin{eqnarray}
C_{0}&=&\frac{\alpha_{g}}{8\pi^{3}\delta_{2}^{2}}
\int_{-\infty}^{+\infty}dx\int_{-\infty}^{+\infty}dy
\frac{-x^{2}+1+\delta_{1}^{2}y^{2}}{\left(x^{2}+1 +
\delta_{1}^{2}y^{2}\right)^{2}}
\frac{1}{\sqrt{\frac{1}{\delta_{2}^{2}}x^{2}+1 +
\frac{\delta_{3}^{2}}{\delta_{2}^{2}}y^{2}}},
\\
C_{1}&=&\frac{\alpha_{g}}{8\pi^{3}\delta_{2}^{2}}
\int_{-\infty}^{+\infty}dx\int_{-\infty}^{+\infty}dy
\frac{x^{2}-1+\delta_{1}^{2}y^{2}}{\left(x^{2}+1+\delta_{1}^{2}
y^{2}\right)^{2}} \frac{1}{\sqrt{\frac{1}{\delta_{2}^{2}}x^{2}+1 +
\frac{\delta_{3}^{2}}{\delta_{2}^{2}}y^{2}}},
\\
C_{2}&=&\frac{\alpha_{g}}{8\pi^{3}\delta_{2}^{2}}
\int_{-\infty}^{+\infty}dx\int_{-\infty}^{+\infty}dy
\frac{x^{2}+1-\delta_{1}^{2}y^{2}}{\left(x^{2}+1 +
\delta_{1}^{2}y^{2}\right)^{2}}\frac{1}{\sqrt{\frac{1}{\delta_{2}^{2}}
x^{2}+1 +\frac{\delta_{3}^{2}}{\delta_{2}^{2}}y^{2}}},
\\
C_{3}&=&\frac{\alpha_{g}}{8\pi^{3}\delta_{2}^{2}}
\int_{-\infty}^{+\infty}dx\int_{-\infty}^{+\infty}dy
\frac{1}{x^{2}+1+\delta_{1}^{2}y^{2}}
\frac{1}{\sqrt{\frac{1}{\delta_{2}^{2}}x^{2}
+1+\frac{\delta_{3}^{2}}{\delta_{2}^{2}}y^{2}}}, \\
C_{\phi}&=&\frac{Ng^{2}k_{F}}{8\pi
v_{F}v_{z}\Lambda}=\frac{\beta_{g}}{\delta_{1}}, \\
C_{\bot}&=&\frac{g^{2}k_{F}v_{F}}{32\pi v_{z}v_{b\bot}^{2}\Lambda} =
\frac{\beta_{g}}{4\delta_{1}\delta_{2}^{2}}, \\
C_{z}&=&\frac{g^{2}k_{F}v_{z}}{16\pi v_{F}v_{bz}^{2}\Lambda} =
\frac{\beta_{g}\delta_{1}}{2\delta_{3}^{2}}.
\end{eqnarray}

\begin{figure}[htbp]
\center
\includegraphics[width=3.38in]{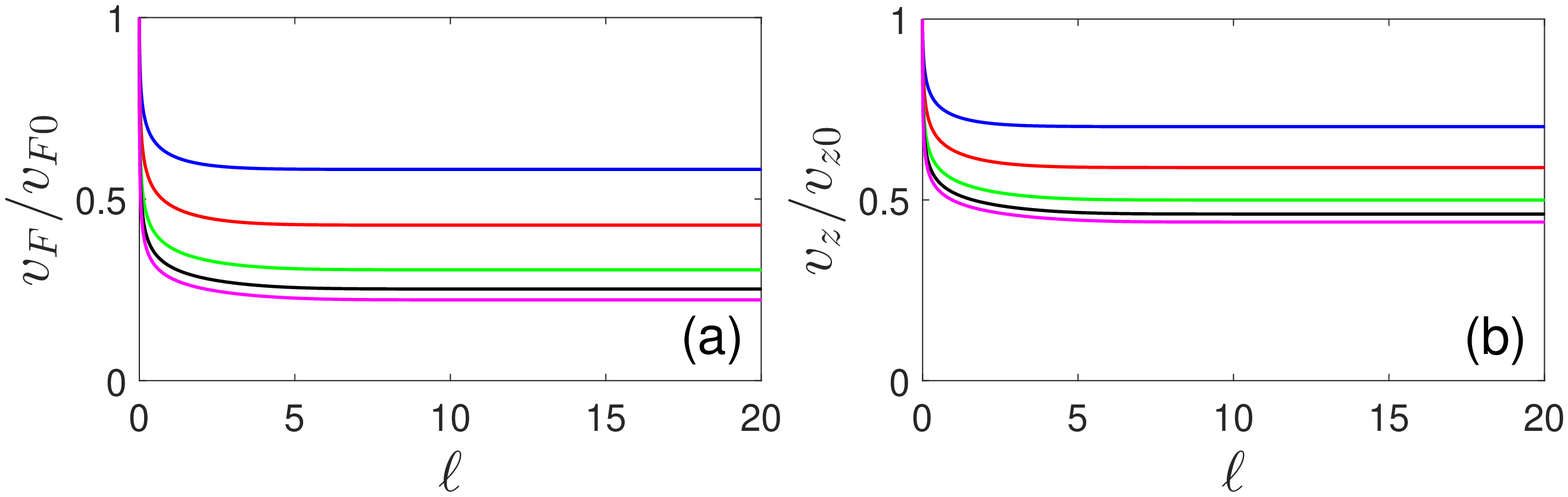}
\includegraphics[width=3.38in]{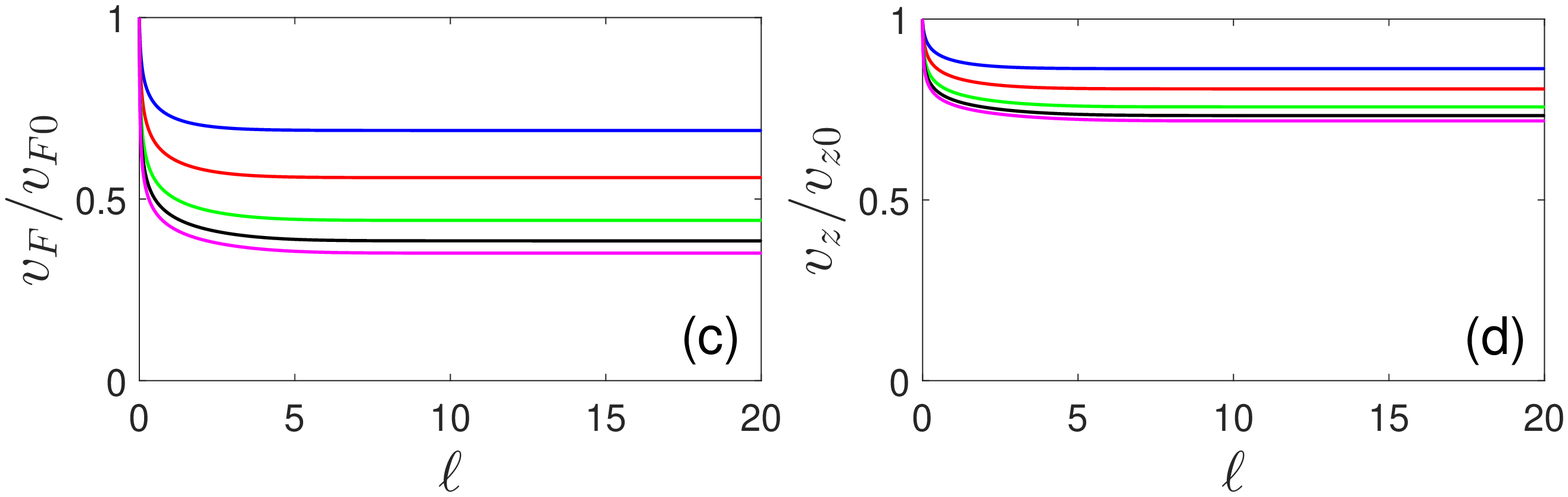}
\includegraphics[width=3.38in]{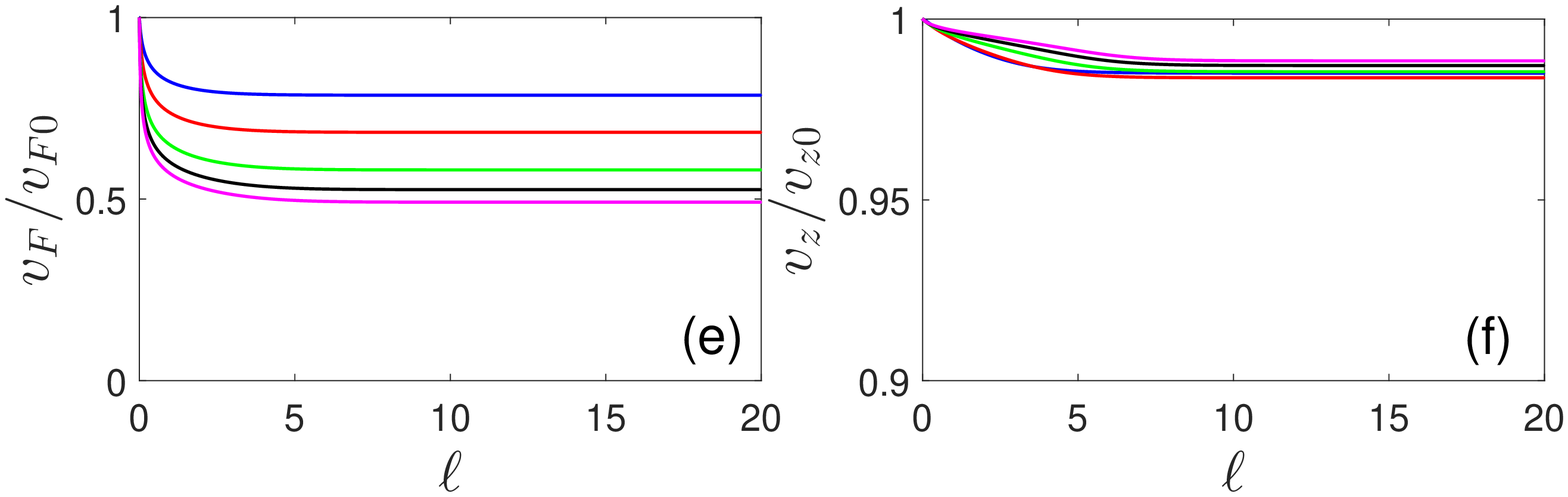}
\includegraphics[width=3.38in]{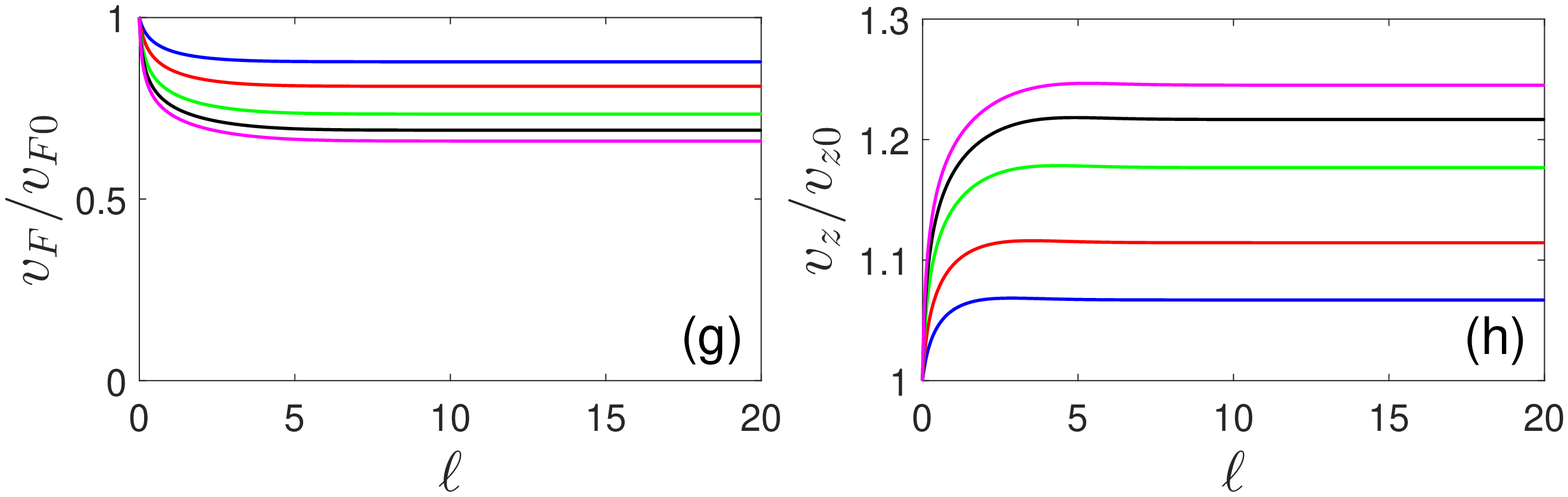}
\includegraphics[width=3.38in]{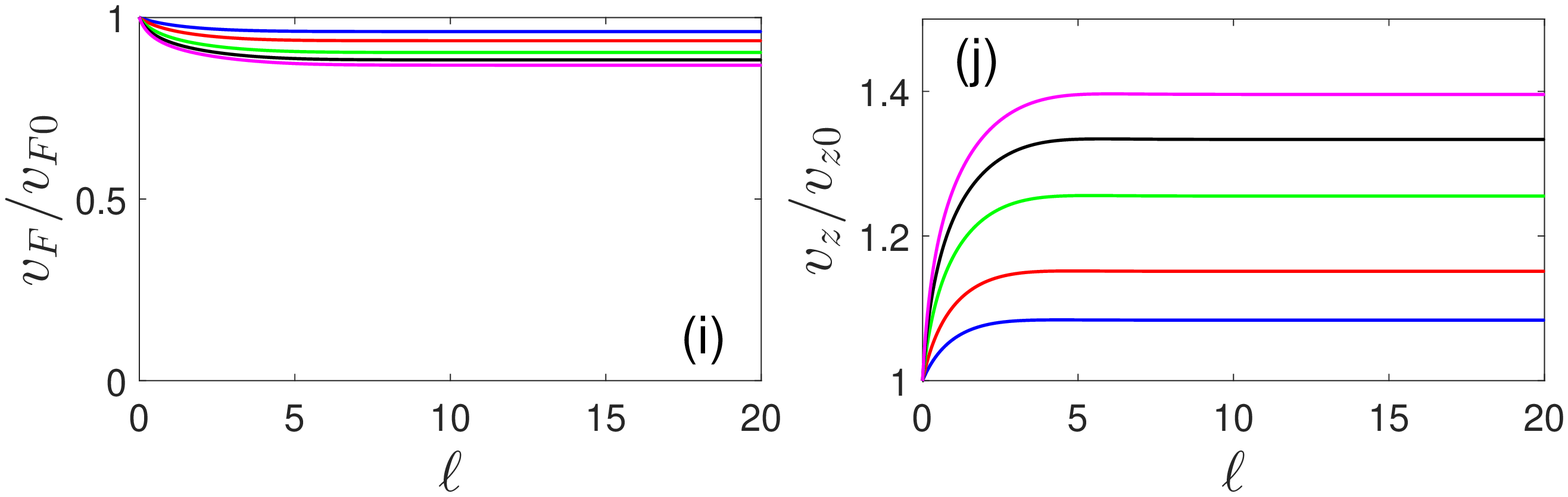}
\includegraphics[width=3.38in]{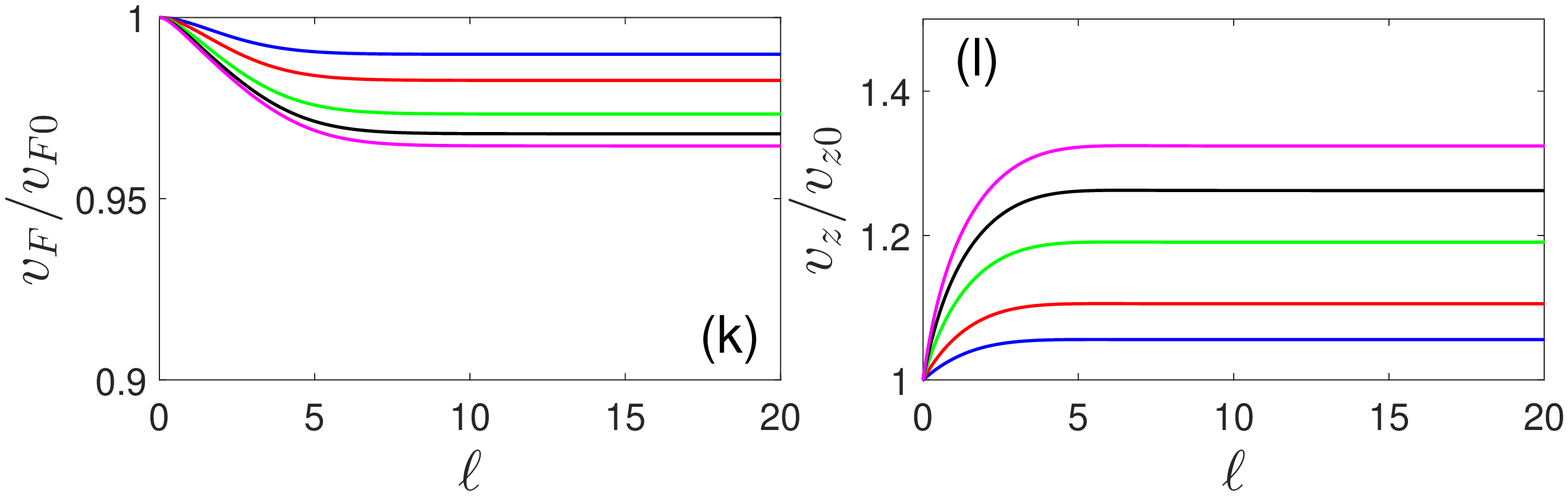}
\includegraphics[width=3.38in]{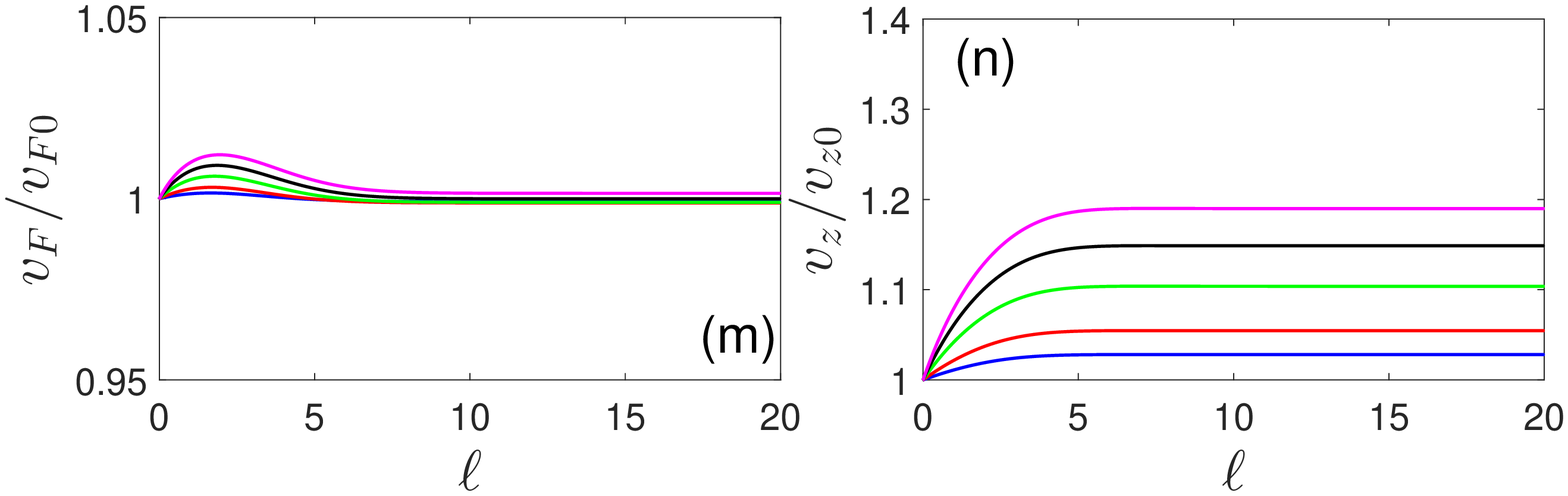}
\includegraphics[width=3.38in]{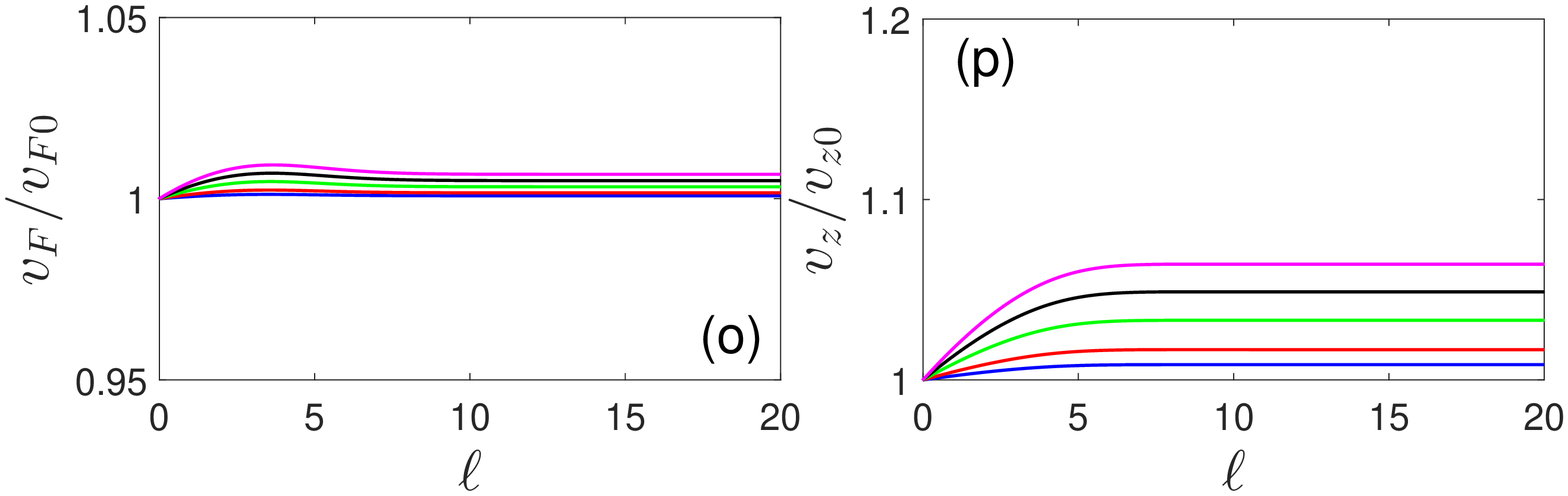}
\caption{Flows of $v_{F}$ and $v_{z}$ for different initial
conditions. (a) and (b): $\delta_{20}=\delta_{30}=0.02$; (c) and
(d): $\delta_{20}=\delta_{30}=0.05$; (e) and (f):
$\delta_{20}=\delta_{30}=0.1$; (g) and (h):
$\delta_{20}=\delta_{30}=0.2$; (i) and (j):
$\delta_{20}=\delta_{30}=0.5$; (k) and (l):
$\delta_{20}=\delta_{30}=1$;  (m) and (n):
$\delta_{20}=\delta_{30}=2$; (o) and (p):
$\delta_{20}=\delta_{30}=5$. Blue, red, green, black, magenta curves
represent the initial values $\alpha_{g0}=1, 2, 4, 6, 8$. The
initial conditions $\delta_{10}=0.1$, $\beta_{g0}=0.1$ and $u_{0}=1$
are always taken. \label{Fig:VRGFermionDiffRatioBF}}
\end{figure}

The RG flows of fermion velocities $v_{F}$ and $v_{z}$ for different
initial conditions are shown in
Fig.~\ref{Fig:VRGFermionDiffRatioBF}. To make our analysis more
generic, we let the ratios $\delta_{20}$ and $\delta_{30}$ to vary
within a wide range. We see that, for small values of $\delta_{20}$
and $\delta_{30}$ (for simplicity $\delta_{20}$ and $\delta_{30}$
are assumed to be equal), $v_{F}$ and $v_{z}$ are both obviously
suppressed. Accordingly, the effective fermion mass is enhanced.
However, for sufficiently large values of $\delta_{20}$ and
$\delta_{30}$, corresponding to the case in which nodal line
fermions move more slowly than the boson (excitonic fluctuation),
$v_{F}$ and $v_{z}$ are slightly increased. Therefore, in order to
obtain the observed fermion mass enhancement in ZrSiS, the bare
(unrenormalized) velocity of the boson mode (excitonic fluctuation)
ought to be smaller than the bare velocity of nodal line fermions.

If the system is in the SM side of the QCP, the parameter $r$ would
take a finite value. For finite $r$, the propagator of boson
describing the excitonic fluctuation has the form
\begin{eqnarray}
D_{0}(\Omega,\mathbf{q}) =
\frac{1}{\Omega^{2}+v_{b\bot}^{2}q_{\bot}^{2} +
v_{bz}^{2}q_{z}^{2}+r}. \label{Eq:BosonPropagatorFiniter}
\end{eqnarray}
It is clear that, the excitonic quantum fluctuation is suppressed by
finite $r$, since $D_{0}(\Omega,\mathbf{q})$ becomes smaller. As $r$
continuously increases from zero, the system is tuned to depart from
the QCP into the SM phase. In this process, the mass enhancement is
gradually weakened with growing $r$, and finally disappears at
sufficiently large $r$. As long as the system is not far from the
QCP, the quasiparticle mass is always enhanced comparing to the bare
mass.


\section{Influence of long-range Coulomb interaction \label{Sec:DerivationCoulomb}}

Here we provide the calculational details of the RG equations caused
by long-range Coulomb interaction. The results show that Coulomb
interaction tends to reduce the quasiparticle mass, inconsistent
with the observed mass enhancement in ZrSiS.

\subsection{Model}

The influence of Coulomb interaction on the nodal line fermions can
be described by the coupling between the fermion field $\psi$ and a
boson field $\phi_{e}$ through
\begin{eqnarray}
S_{\psi\phi_{e}} = ie\int\frac{d\omega}{2\pi}
\frac{d^3\mathbf{k}}{(2\pi)^{3}}\frac{d\Omega}{2\pi}
\frac{d^3\mathbf{q}}{(2\pi)^{3}}\psi^{\dag}(\omega,\mathbf{k})
\psi(\omega+\Omega,\mathbf{k}+\mathbf{q})\phi_{e}(\Omega,\mathbf{q}),
\end{eqnarray}
which can be further written as
\begin{eqnarray}
S_{\psi\phi_{e}} = ie k_{F}\int\frac{d\omega}{2\pi}
\frac{dk_{r}}{2\pi}\frac{dk_{z}}{2\pi}\frac{d\Omega}{2\pi}
\frac{d^3\mathbf{q}}{(2\pi)^{3}}\psi^{\dag}(\omega,\mathbf{k})
\psi(\omega+\Omega,\mathbf{k}+\mathbf{q})\phi_{e}(\Omega,\mathbf{q}).
\end{eqnarray}
The action for the fermion field $\psi$ still can be written as
Eq.~(\ref{Eq:ActionNodalFermions}). The action for the boson field
$\phi_{e}$ takes the form
\begin{eqnarray}
S_{\phi_{e}}=\int\frac{d\Omega}{2\pi}\frac{dq_{x}}{2\pi}
\frac{dq_{y}}{2\pi}\frac{dq_{z}}{2\pi}\phi_{e}(\Omega,\mathbf{q})
\left(a\left(q_{x}^{2}+q_{y}^{2}\right)+\frac{1}{a} q_{z}^{2}
\right)\phi_{e}(\Omega,\mathbf{q}).
\end{eqnarray}

The propagator of boson $\phi_{e}$ is given by
\begin{eqnarray}
D_{e0}(\mathbf{q})=\frac{1}{a\left(q_{x}^{2}+q_{y}^{2}\right)+\frac{1}{a}q_{z}^{2}}.
\label{Eq:BosonCoulombPropagator}
\end{eqnarray}

\subsection{The self-energy of boson}

The self-energy of boson is defined as
\begin{eqnarray}
\Pi_{e}(\Omega,\mathbf{q})&=&e^{2}\int'\frac{d\omega}{2\pi}
\frac{d^3\mathbf{k}}{(2\pi)^{3}}
\mathrm{Tr}\left[G_{0}\left(\omega,\mathbf{k}\right)
G_{0}\left(\omega+\Omega,\mathbf{k}+\mathbf{q}\right) \right].
\label{Eq:PolaCoulombDef}
\end{eqnarray}
Substituting Eq.~(\ref{Eq:FermionPropagator}) into
Eq.~(\ref{Eq:PolaCoulombDef}) and expanding to the quadratic order
of $q_{r}$ and $q_{z}$, we obtain that the self-energy of boson in
the zero energy limit takes the form
\begin{eqnarray}
\Pi_{e}(\mathbf{q})
&=&-v_{F}^{2}q_{r}^{2}\frac{e^{2}}{4}\int'\frac{d^3\mathbf{k}}{(2\pi)^{3}}
\frac{v_{z}^{2}k_{z}^{2}}{E_{\mathbf{k}}^{5}}\cos^{2}(\theta)
-v_{z}^{2}q_{z}^{2}\frac{e^{2}}{4}\int'\frac{d^3\mathbf{k}}{(2\pi)^{3}}
\frac{v_{F}^{2}k_{r}^{2}}{E_{\mathbf{k}}^{5}}.
\end{eqnarray}
Carrying out the integration of momenta by adopting the RG scheme
shown in Eq.~(\ref{Eq:RGSchemeFermion}), we arrive at
\begin{eqnarray}
\Pi_{e}(\mathbf{q})
&=&-C_{e\bot}q_{\bot}^{2}\ell-C_{ez}q_{z}^{2}\ell,
\end{eqnarray}
where
\begin{eqnarray}
C_{e\bot}=\frac{\beta_{e}}{\sqrt{2}\delta_{1}},\qquad
C_{ez}=\beta_{e}\sqrt{2}\delta_{1},
\end{eqnarray}
with
\begin{eqnarray}
\beta_{e}=\frac{\sqrt{2}e^{2}k_{F}}{32\pi \Lambda},\qquad
\delta_{1}=\frac{v_{z}}{v_{F}}.
\end{eqnarray}

\subsection{The self-energy of fermions}

The self-energy of fermions induced by Coulomb interaction reads as
\begin{eqnarray}
\Sigma_{C}(\omega,k_{F}+\mathbf{k}) = -e^{2}\int'
\frac{d^3\mathbf{q}}{(2\pi)^{3}} \frac{d\Omega}{2\pi}
G_{0}(\mathbf{k}+\mathbf{q},i\omega+i\Omega)D_{e0}(\mathbf{q}).
\label{Eq:SelfEnergyFermionCoulombDef}
\end{eqnarray}
Substituting Eqs.~(\ref{Eq:FermionPropagator}) and
(\ref{Eq:BosonCoulombPropagator}) into
Eq.~(\ref{Eq:SelfEnergyFermionCoulombDef}) and performing the
integration of $\Omega$ gives rise to
\begin{eqnarray}
\Sigma_{C}(\omega,k_{F}+\mathbf{k})&=&-\frac{e^{2}}{2}\int'
\frac{d^3\mathbf{q}}{(2\pi)^{3}}\frac{v_{F}\left(k_{x}+q_{x}\right)\sigma_{1}
+v_{z}\left(k_{z}+q_{z}\right)\sigma_{2}}{\sqrt{v_{F}^{2}\left(k_{x}+q_{x}\right)^{2}
+v_{z}^{2}\left(k_{z}+q_{z}\right)^{2}}}\frac{1}{a\left(q_{x}^{2}+q_{y}^{2}\right)
+ \frac{1}{a}q_{z}^{2}}.
\end{eqnarray}
Expanding to the leading order of $\mathbf{k}_{i}$, we obtain
\begin{eqnarray}
\Sigma_{C}(\omega,k_{F}+\mathbf{k}) &=&
-v_{F}k_{x}\sigma_{1}\frac{e^{2}}{2}\int'\frac{d^3\mathbf{q}}{(2\pi)^{3}}
\frac{v_{z}^{2}q_{z}^{2}}{\left(v_{F}^{2}q_{x}^{2} +
v_{z}^{2}q_{z}^{2}\right)^{\frac{3}{2}}}\frac{1}{a\left(q_{x}^{2}+q_{y}^{2}\right)
+ \frac{1}{a}q_{z}^{2}}\nonumber
\\
&&-v_{z}k_{z}\sigma_{2}\frac{e^{2}}{2}\int'\frac{d^3\mathbf{q}}{(2\pi)^{3}}
\frac{v_{F}^{2}q_{x}^{2}}{\left(v_{F}^{2}q_{x}^{2} +
v_{z}^{2}q_{z}^{2}\right)^{\frac{3}{2}}}\frac{1}{a\left(q_{x}^{2} +
q_{y}^{2}\right)+\frac{1}{a}q_{z}^{2}}.
\end{eqnarray}
Utilizing the RG scheme
\begin{eqnarray}
\int'\frac{d^3\mathbf{q}}{(2\pi)^{3}}=\frac{1}{8\pi^{3}}
\left(\int_{b\Lambda}^{\Lambda}+\int_{-\Lambda}^{-b\Lambda}\right)dq_{x}
\int_{-\infty}^{+\infty}dq_{y}\int_{-\infty}^{+\infty}dq_{z},
\end{eqnarray}
we can get
\begin{eqnarray}
\Sigma_{C}(\omega,k_{F}+\mathbf{k})
&=&-v_{F}k_{x}\sigma_{x}C_{e1}\ell -v_{z}k_{z}\sigma_{y}C_{e2}\ell,
\end{eqnarray}
where
\begin{eqnarray}
C_{e1}=\frac{\alpha_{e}}{8\pi^{2}}F_{1}(a\delta_{1}),\qquad
C_{e2}=\frac{\alpha_{e}}{8\pi^{2}} F_{2}(a\delta_{1}),
\end{eqnarray}
with
\begin{eqnarray}
\alpha_{e}&=&\frac{e^{2}}{v_{F}},
\\
F_{1}(a\delta_{1})&=&\int_{-\infty}^{+\infty}dz\frac{a^{2}
\delta_{1}^{2}z^{2}}{\left(1+a^{2}\delta_{1}^{2}z^{2}\right)^{\frac{3}{2}}}
\frac{1}{\sqrt{1+z^{2}}}, \\
F_{2}(a\delta_{1})&=&\int_{-\infty}^{+\infty}dz\frac{1}{\left(1 +
a^{2}\delta_{1}^{2}z^{2}\right)^{\frac{3}{2}}}
\frac{1}{\sqrt{1+z^{2}}}.
\end{eqnarray}

\subsection{Vertex Correction}

The correction for the fermion-boson coupling is defined as
\begin{eqnarray}
\Gamma_{e}=e^{2}\int'\frac{d\Omega}{2\pi}\frac{d^3\mathbf{q}}{(2\pi)^{3}}
G_{0}(\Omega,\mathbf{q})G_{0}(\Omega,\mathbf{q})
D_{e0}(\mathbf{q}). \label{Eq:VortexCorrectionCoulomb}
\end{eqnarray}
Substituting Eqs.~(\ref{Eq:FermionPropagator}) and
(\ref{Eq:BosonCoulombPropagator}) into
Eq.~(\ref{Eq:VortexCorrectionCoulomb}), we find that
\begin{eqnarray}
\Gamma_{e}=0.
\end{eqnarray}

\subsection{Derivation of the RG equations }

The free action of $\psi$ is
\begin{eqnarray}
S_{\psi} = \int\frac{d\omega}{2\pi}\frac{dk_{x}}{2\pi}
\frac{dk_{y}}{2\pi}\frac{dk_{z}}{2\pi}\psi^{\dag}(\omega,\mathbf{k})
\left(-i\omega+\frac{k_{x}^{2}+k_{y}^{2}-k_{F}^{2}}{2m}\sigma_{x}
+v_{z}k_{z}\sigma_{y} \right)\psi(\omega,\mathbf{k}).
\end{eqnarray}
In the low-energy regime, the free action of $\psi$ can be also
written as
\begin{eqnarray}
S_{\psi}=k_{F}\int\frac{d\omega}{2\pi}\frac{dk_{r}}{2\pi}
\frac{dk_{z}}{2\pi}\psi^{\dag}(\omega,\mathbf{k})
\left(-i\omega+v_{F}k_{r}\sigma_{x} +v_{z}k_{z}\sigma_{y}
\right)\psi(\omega,\mathbf{k}).
\end{eqnarray}
Considering the correction of self-energy of the fermions, the
action becomes
\begin{eqnarray}
S_{\psi}&=&k_{F}\int\frac{d\omega}{2\pi}\frac{dk_{r}}{2\pi}
\frac{dk_{z}}{2\pi}\psi^{\dag}(\omega,\mathbf{k})
\left(-i\omega+v_{F}k_{r}\sigma_{x} + v_{z}k_{z}\sigma_{y} -
\Sigma_{C}(\omega,\mathbf{k}) \right)\psi(\omega,\mathbf{k})\nonumber
\\
&\approx&k_{F}\int\frac{d\omega}{2\pi}\frac{dk_{r}}{2\pi}
\frac{dk_{z}}{2\pi}\psi^{\dag}(\omega,\mathbf{k})
\left(-i\omega+v_{F}k_{r}\sigma_{x}e^{C_{e1}\ell} + v_{z}k_{z}
\sigma_{y}e^{C_{e2}\ell} \right)\psi(\omega,\mathbf{k}).
\end{eqnarray}
Utilizing the transformations
\begin{eqnarray}
\omega&=&\omega'e^{-\ell}  \label{Eq:Coulombomegascaling}
\\
k_{r}&=&k_{r}'e^{-\ell}, \label{Eq:Coulombkrscaling}
\\
k_{z}&=&k_{z}'e^{-\ell},  \label{Eq:Coulombkzscaling}
\\
\psi&=&\psi' e^{2\ell},  \label{Eq:Coulombpsiscaling}
\\
v_{F}&=&v_{F}'e^{-C_{e1}\ell},  \label{Eq:CoulombvFscaling}
\\
v_{z}&=&v_{z}'e^{-C_{e2}\ell},  \label{Eq:Coulombvzscaling}
\end{eqnarray}
the action becomes
\begin{eqnarray}
S_{\psi}=k_{F}\int\frac{d\omega'}{2\pi}\frac{dk_{r}'}{2\pi}
\frac{dk_{z}'}{2\pi}\psi'^{\dag}(\omega',\mathbf{k}')
\left(-i\omega'+v_{F}k_{r}'\sigma_{x} +v_{z}k_{z}'\sigma_{y}
\right)\psi'(\omega',\mathbf{k}'),
\end{eqnarray}
which recovers the original form of the actin.

The free action of the boson field $\phi_{e}$ is
\begin{eqnarray}
S_{\phi_{e}}=\int\frac{d\Omega}{2\pi}\frac{dq_{x}}{2\pi}
\frac{dq_{y}}{2\pi}\frac{dq_{z}}{2\pi}\phi_{e}(\Omega,\mathbf{q})
\left(a\left(q_{x}^{2}+q_{y}^{2}\right)+\frac{1}{a} q_{z}^{2}
\right)\phi_{e}(\Omega,\mathbf{q}).
\end{eqnarray}
Including the correction of self-energy of boson, the action of
$\phi_{e}$ becomes
\begin{eqnarray}
S_{\phi_{e}}&=&\int\frac{d\Omega}{2\pi}\frac{dq_{x}}{2\pi}
\frac{dq_{y}}{2\pi}\frac{dq_{z}}{2\pi}\phi_{e}(\Omega,\mathbf{q})
\left(aq_{r}^{2}+\frac{1}{a} q_{z}^{2}-\Pi_{e}(\mathbf{q},0)
\right)\phi_{e}(\Omega,\mathbf{q})\nonumber
\\
&\approx&\int\frac{d\Omega}{2\pi}\frac{dq_{x}}{2\pi}
\frac{dq_{y}}{2\pi}\frac{dq_{z}}{2\pi}\phi_{e}(\Omega,\mathbf{q})
\left(aq_{r}^{2}e^{\frac{C_{\bot}}{a}\ell}+\frac{1}{a}
q_{z}^{2}e^{aC_{z}\ell} \right)\phi_{e}(\Omega,\mathbf{q}).
\end{eqnarray}
Using the scaling transformations
\begin{eqnarray}
\Omega&=&\Omega'e^{-\ell}, \label{Eq:CoulombOmegaLscaling}
\\
q_{x}&=&q_{x}'e^{-\ell}, \label{Eq:Coulombqxscaling}
\\
q_{y}&=&q_{y}'e^{-\ell}, \label{Eq:Coulombqysaling}
\\
q_{z}&=&q_{z}'e^{-\ell}, \label{Eq:Coulombqzcaling}
\\
\phi_{e}&=&\phi_{e}'e^{\left[3-\frac{1}{4}\left(\frac{C_{e\bot}}{a}
+ aC_{ez}\right)\right]\ell}, \label{Eq:Coulombphiscaling}
\\
a&=&a'e^{\left[\frac{1}{2}\left(aC_{ez}-\frac{C_{e\bot}}{a}\right)\right]\ell},
\label{Eq:Coulombascaling}
\end{eqnarray}
the action of $\phi_{e}'$ becomes
\begin{eqnarray}
S_{\phi_{e}'} = \int\frac{d\Omega'}{2\pi}\frac{dq_{x}'}{2\pi}
\frac{dq_{y}'}{2\pi}\frac{dq_{z}'}{2\pi}\phi_{e}'(\Omega',\mathbf{q}')
\left(a'q_{r}'^{2}+\frac{1}{a'} q_{z}'^{2}\right)
\phi_{e}'(\Omega',\mathbf{q}'),
\end{eqnarray}
which has the some form as the free action of boson field.

The action of fermion-boson coupling is
\begin{eqnarray}
S_{\psi\phi_{e}} = ie\int\frac{d\omega}{2\pi}
\frac{d^3\mathbf{k}}{(2\pi)^{3}}\frac{d\Omega}{2\pi}
\frac{d^3\mathbf{q}}{(2\pi)^{3}}\psi^{\dag}(\omega,\mathbf{k})
\psi(\omega+\Omega,\mathbf{k}+\mathbf{q})\phi_{e}(\Omega,\mathbf{q}).
\end{eqnarray}
It can be further written as
\begin{eqnarray}
S_{\psi\phi_{e}}=i ek_{F}\int\frac{d\omega}{2\pi}\frac{dk_{r}}{2\pi}
\frac{dk_{z}}{2\pi}\frac{d\Omega}{2\pi}
\frac{d^3\mathbf{q}}{(2\pi)^{3}}\psi^{\dag}(\omega,\mathbf{k})
\psi(\omega+\Omega,\mathbf{k}+\mathbf{q})\phi_{e}(\Omega,\mathbf{q}).
\end{eqnarray}
Since the vertex correction vanishes, the action still takes the form
\begin{eqnarray}
S_{\psi\phi_{e}}= iek_{F}\int\frac{d\omega}{2\pi}\frac{dk_{r}}{2\pi}
\frac{dk_{z}}{2\pi}\frac{d\Omega}{2\pi}
\frac{d^3\mathbf{q}}{(2\pi)^{3}}\psi^{\dag}(\omega,\mathbf{k})
\psi(\omega+\Omega,\mathbf{k}+\mathbf{q})\phi_{e}(\Omega,\mathbf{q}).
\end{eqnarray}
Employing the transformations as shown in
Eqs.~(\ref{Eq:Coulombomegascaling})-(\ref{Eq:Coulombpsiscaling}),
and Eqs.(\ref{Eq:CoulombOmegaLscaling})-(\ref{Eq:Coulombascaling}),
and
\begin{eqnarray}
e' = ee^{-\frac{1}{4}\left(\frac{C_{e\bot}}{a}+aC_{ez}\right)\ell},
\label{Eq:Coulombescaling}
\end{eqnarray}
the action becomes
\begin{eqnarray}
S_{\psi'\phi_{e}'} = ie'k_{F}\int\frac{d\omega'}{2\pi}
\frac{dk_{r}'}{2\pi}\frac{dk_{z}'}{2\pi}\frac{d\Omega'}{2\pi}
\frac{d^3\mathbf{q}'}{(2\pi)^{3}}\psi'^{\dag}(\omega',\mathbf{k}')
\psi'(\omega'+\Omega',\mathbf{k}'+\mathbf{q}')\phi_{e}'(\Omega',\mathbf{q}'),
\end{eqnarray}
which recovers the form of the original action.

Form Eqs.~(\ref{Eq:CoulombvFscaling}), (\ref{Eq:Coulombvzscaling}),
(\ref{Eq:Coulombascaling}), (\ref{Eq:Coulombescaling}), we get the
RG equations
\begin{eqnarray}
\frac{dv_{F}}{d\ell}&=&\frac{\alpha}{8\pi^{2}}F_{1}(a\delta_{1})v_{F},
\\
\frac{dv_{z}}{d\ell}&=&\frac{\alpha}{8\pi^{2}} F_{2}(a\delta_{1})v_{z},
\\
\frac{da}{d\ell}&=&\frac{\beta_{e}}{2}\left(\frac{1}{\sqrt{2}a\delta_{1}}
- \sqrt{2}a\delta_{1}\right)a,
\\
\frac{de}{d\ell}&=&-\frac{\beta_{e}}{4}\left(\frac{1}{\sqrt{2}a\delta_{1}}
+ \sqrt{2}a\delta_{1}\right)e,
\\
\frac{d\alpha}{d\ell} &=& \left[-\frac{\beta_{e}}{2}
\left(\frac{1}{\sqrt{2}a\delta_{1}}+\sqrt{2}a\delta_{1}\right)
-\frac{\alpha}{8\pi^{2}}F_{1}(a\delta_{1})\right]\alpha,
\\
\frac{d\beta_{e}}{d\ell} &=& \left[1-\frac{\beta_{e}}{2}
\left(\frac{1}{\sqrt{2}a\delta_{1}}+\sqrt{2}a\delta_{1}\right)\right]
\beta_{e}, \\
\frac{d(a\delta_{1})}{d\ell} &=& \left[\frac{\beta_{e}}{2}
\left(\frac{1}{\sqrt{2}a\delta_{1}}-\sqrt{2}a\delta_{1}\right) +
\frac{\alpha}{8\pi^{2}} \left(F_{2}(a\delta_{1}) -
F_{1}(a\delta_{1})\right)\right]a\delta_{1}.
\end{eqnarray}
These RG equations are very close to the ones presented in
Refs.~\cite{Huh16} and \cite{Wang17}. The minor differences result
from the fact that a different RG scheme is adopted in our
derivation from Refs.~\cite{Huh16} and \cite{Wang17}.

\end{document}